\documentclass[twocolumn,preprintnumbers,amsmath,amssymb,prb]{revtex4}
\pdfoutput=1
\usepackage{graphicx,amsmath}
\usepackage{dcolumn}
\usepackage{bm}
\usepackage{amssymb}
\usepackage{epstopdf}
\usepackage{color}

\begin{document}

\title{Formation of probability density waves and probability current density waves by excitation and decay of a doublet of quasistationary states of a three-barrier heterostructure upon scattering of gaussian wave packets}

\begin{abstract}
Annotation. A numerical-analytical simulation of scattering by a three-barrier heterostructure of an electronic Gaussian wave packet, the spectral width of which is on the order of the distance between the levels of the doublet of quasi-stationary states, is carried out. It is shown that as a result of scattering, damped waves of electron charge and current densities are formed outside the double well, their characteristics are determined by the structure of the initial wave packet and the poles of the scattering amplitudes. The frequency of these waves is equal to the difference frequency of the doublet, the wavenumber is the difference between the wave numbers of free motion of electrons with resonant energies, and the speed of their propagation is the ratio of these quantities. The system can go into the regime of repetition or amplification of the emission of electron waves if a periodic resonant pumping of the doublet population is provided by scattering of a series of coherent wave packets.
\end{abstract}

\pacs{84.40.Az,~ 84.40.Dc,~ 85.25.Hv,~ 42.50.Dv,~42.50.Pq}
 \keywords      { quantum }

\date{\today }

\author{Yu. G. Peisakhovich}
\affiliation{Novosibirsk State Technical University, Novosibirsk,
Russia}
\author{A. A. Shtygashev} \email{shtygashev@corp.nstu.ru}
\affiliation{Novosibirsk State
Technical University, Novosibirsk, Russia}

%\classification{85.25.Dq,~ 85.25.Cp,~ 85.25.Hv,~ 84.40.Az}

 \maketitle

\section{Introduction}\label{intr}
The ability of nanoheterostructures to selectively transmit and convert wave signals of different physical nature makes it possible to create high-speed and high-frequency devices for optoelectronics, acoustoelectronics, information transmission systems, laser technology, etc. In recent decades, laser light sources have been created capable of generating ultrashort pulses of picosecond, femtosecond, and even attosecond duration \cite{Rost2011}-\cite{Chek2014}. This stimulated the intensive development of spectroscopy and high technologies in the corresponding frequency ranges. The impact of such short-term signals on microscopic and macroscopic systems and the detection of responses make it possible to study fast processes, the duration of which is less than or on the order of the relaxation times in the systems \cite{Rost2011}-\cite{Ross2002}. In addition to spectroscopic sensing of matter, it is possible to pose the problem of generating an alternating current in the terahertz range by converting ultrashort excitation pulses into a system of oscillations and waves of electron density of charge and current on scales smaller than the length and time of quantum coherence of electrons. This problem can be solved using nanoscale heterostructures. It is well known that in thin-film nanostructures such as a double quantum well with tunnel-transparent walls for electrons, the energy spectrum of the transverse motion of electrons contains doublets of resonance levels that are relatively close to each other. In the forbidden bands of film below the vacuum level, such a spectrum is discrete and the wave functions of doublet states are localized in the well. In the allowed bands below and above the vacuum level, the energy spectrum is continuous and the wave functions of resonance doublets describe delocalized quasi-stationary states of the transverse scattering problem. The energies of the doublets and the lifetimes of quasi-stationary states are determined by the poles of the amplitudes of stationary electron scattering by the heterostructure, as well as by the shift and smearing of levels due to inelastic electron scattering. Pulsed excitation and slow decay of a quasi-resonant nonstationary state formed by the superposition and interference of quantum states from a narrow band of the electronic spectrum that includes a doublet can be accompanied by beats of the space-time distributions of the probability densities and current of electrons whose energies belong to such a narrow band. This kind of beating often accompany a quantum transient \cite{Leo1991}-\cite{Cald2016} after a single pulse excitation and last for the lifetime of quasi-stationary states, which can be much longer than the time period of these beats if the transparency of the barriers is sufficiently low and the inelastic processes for electrons are weak. This effect was first observed indirectly in experiments on differential transmission and four-wave mixing for femtosecond light pulses in an asymmetric double quantum well \cite{Leo1991}-\cite{Rosk1992}. In such a well, the quantum beats of the superposition of the wave functions of the doublet of stationary states of the discrete spectrum of transverse motion cause the appearance of resonant damped oscillations of the electron-hole dipole moment and a certain number of registered oscillations of the dipole electromagnetic radiation at the terahertz difference frequency of the doublet.

A similar effect should also exist in the case when the doublet of quasi-stationary states of the transverse scattering problem is located in the continuous spectrum of the conduction bands above or below the vacuum level \cite{Romo2002}-\cite{Peis2008B}. In this paper, it will be shown that if the transparency of the potential barriers of the heterostructure is sufficiently low, then the coupled oscillations of mixed doublet resonance states should manifest themselves not only in the periodic flow of the electron density between the wells through the middle barrier inside the double well \cite{Peis2008A}-\cite{Cald2011}, but they should also be accompanied by oscillations of the charge density and current electrons escaping into outer space through extreme potential barriers. Outside the double well, these spatiotemporal oscillations of the envelopes of the charge and current densities can have the character of waves traveling to the left and right from the heterostructure and decaying in time and space. The frequency of these waves is equal to the difference frequency of the doublet, the wavenumber is the difference between the wave numbers of free motion of electrons with resonant energies, and the speed of their propagation is the ratio of these quantities. The process of emission of such electron waves lasts for the lifetime of quasi-stationary states, which can be much longer than the wave period if the transparency of the barriers is sufficiently low. With distance from the heterostructure, trains of difference waves of charge and current densities decay and broaden rather slowly and can be detected and removed from the system using electric and magnetic fields of the corresponding structure. The system can switch to the mode of repetition of the emission of electron waves or even to the mode of 
self-oscillation if positive feedback and periodic resonant pumping of the population of the doublet in the heterostructure are provided.

Population and decay of a doublet of quasistationary states can be provided in different ways. We theoretically studied and simulated various mechanisms of this process. The first of them consists in the scattering of a Gaussian electron wave packet incident on a double-well system from the outside. In the leading approximation, it can be described by a relatively simple quantum-mechanical model in the language of only pure one-particle quantum states of the scattering problem, which makes it possible to rigorously reveal the main laws of the process and estimate the contributions of the main features. It turned out that the amplitude of the resonant difference spatio-temporal wave harmonic can be greater than or of the order of the amplitudes of its smooth and high-frequency components. These results are presented below in this paper using the example of a one-dimensional model of scattering of a Gaussian wave packet by a structure with three identical  $\delta$-barriers.

Two other mechanisms of the population and decay of the doublet of quasi-stationary states that we studied are associated with diffraction by a double-well heterostructure of photoelectrons arising from the action of an ultrashort light pulse on the photocathode with the subsequent formation of a kind of alternating photoemission current. One mechanism is provided by the incidence of a photoelectron pulse from the outside onto a double-well heterostructure deposited on a bulk planar photocathode, and the other is provided by pulsed photoexcitation of electrons directly in thin layers from the inside of the double-well structure, which itself acts as a very thin photocathode. To describe these methods of excitation and decay, it is necessary to consider mixed quantum states taking into account the external high-frequency electromagnetic pumping field, as well as inelastic scattering of electrons, using the approximate methods of the nonstationary quantum theory of many bodies. For this we used the mathematical apparatus of the density matrix. The results of an approximate description and calculations will be presented in the following articles, where it will be shown that the population of quasi-stationary levels can be determined not only by the explicit pole features of the amplitudes of resonant scattering of electrons on a double-well heterostructure, but also by the contribution of these features to the photoexcitation spectrum, and even to the magnitude of the matrix elements of electronic optical transitions upon photoexcitation from within the heterostructure. 

Here we are interested in the time and space oscillating solutions of the one-dimensional nonstationary Schrodinger equation with the potential in the form of wells and barriers, which are located in a finite region. To obtain such solutions, three methods are most often used:  a) direct numerical integration in finite differences \cite{Kons2003}, b) calculation of the dynamic superposition of solutions of the stationary Schrodinger equation for a boundary value problem with a continuous and/or discrete spectrum \cite{Peis2008A},\cite{Peis2008B}, \cite{Wint1961},\cite{Cald2013},\cite{Cald2016}, describing the evolution of a wave packet, c) representation of the solution in the form of a resonant expansion, the members of which are the products of the Moshinsky function (associated with the problem of a quantum gate and diffraction in time) and resonant wave functions in the internal region of the potential, where they are finite and normalized by specific conditions. The latter method c) was developed by G. Garcia-Calderon et al. \cite{Camp2009}-\cite{Cald2016}. They carried out active research of transient quantum processes in resonant tunneling and published many articles containing interesting and important results describing the evolution and asymptotics of electronic wave functions in different regions of time and space. Most of these details relate to the internal region of action of the potential, where the form of the resonant wave functions is known. In particular, as in our papers  \cite{Peis2008A}-\cite{Peis2008B}, the impulsive character of the decay of quasi-stationary states was illustrated  \cite{Cald2009} if the spectrum of the initial wave packet covers a small number of quasi-resonant levels; the flow of the wave function between successive wells was called in \cite{Cald2009},\cite{Cald2011} the "bouncing" and "breathing" modes. In \cite{Cald2013},\cite{Cald2016}, general formulas for decay wave functions outside the region of action of the potential were also written, the coordinate dependence of these functions was determined by the Moshinsky functions, and not by the resonance Gamow wave functions, which exponentially increase with increasing distance from the system. However, outside the region of action of the potential, the wave character of the behavior of the probability and current densities during the decay of a mixture of a doublet of quasistationary states of a two-well system, considered in our work, was not clearly distinguished and discussed in \cite{Cald2013},\cite{Cald2016}.

 In contrast, in this article, when describing the scattering of one or a system of Gaussian pulses, we focus on not only the inside, but also on the outside region of action of the double-well potential. We use method b) to describe nonstationary probability densities and probability currents at an arbitrary point in space, and show that in the outer region of a double quantum well, the envelopes of these quantities demonstrate the properties of traveling waves. Calculations by the method of continual decomposition b) are not the calculations of a ""black box" type", that supposedly "provide no deep physical insight" and "does not provide grasp of the time evolution of the initial state" \cite{Cald2007}-\cite{Cald2016}. There is developed by G.F. Drukarev in 1951 \cite{Druk1951},\cite{Baz1969} an elegant version of the saddle-point method, which allows, within the framework of method b), to identify and estimate the oscillating contributions of the pole features of the scattering amplitudes to the wave function of the scattered wave packet in the internal and external regions of the action of the scattering potential, as was done in \cite{Peis2008A}-\cite{Peis2008B}.

At the end of this introduction, we emphasize that the main thing for us here  is that it is the complete system of wave functions of the stationary scattering problem of method b) that provides a natural basis for unperturbed states of the zero approximation for describing and calculating the interactions of electrons with photons and with other particles in the subsequent application of the density matrices method to the problem of photoemission in an open system.

The article is organized as follows. Section II describes a theoretical quantum mechanical model, provides rigorous formulas, and discusses the optimal parameters for describing the scattering of a Gaussian wave packet by a double quantum well. Section III presents the results of rigorous calculations of space-time oscillations and waves of probability densities and currents with an explanation of their characteristics. An approximate method for the analytical identification of these characteristics is described in Appendix A, and the clear but cumbersome expressions for the probability and current densities obtained by this method are given in Appendix B.

\section{THEORETICAL MODEL, CALCULATION FORMULAS AND PARAMETERS}

To confirm the statements made and to highlight the basic laws of the process, in accordance with the algorithm described in our articles \cite{Peis2008A}-\cite{Peis2008B}, we will analyze in detail a simple one-dimensional model, which describes the population and subsequent decay of a doublet of quasi-stationary states of a three-barrier heterostructure due to scattering of pulsed Gaussian wave packets arriving from the left and having a spectral width of the order of the distance between the levels of the doublet. 

\begin{figure}[h]
\includegraphics[width=8.5 cm]{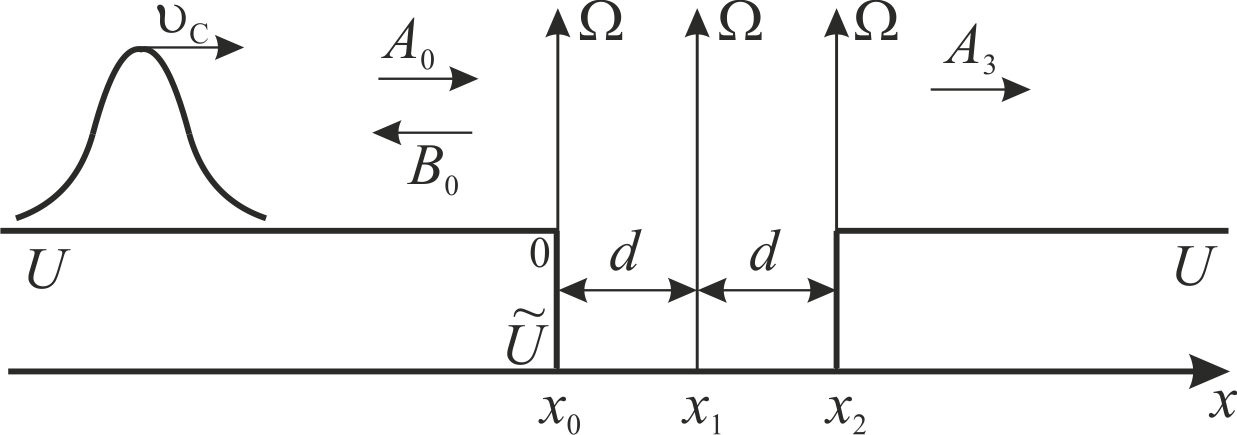}\\
  \caption{Model three-barrier heterostructure. The vertical lines with arrows picture the  $\delta$-barriers.}\label{FIG:Fig1}
\end{figure}

The double quantum well is assumed to be flat, the axis $x$  is directed perpendicular to it, and the origin of coordinates is placed on its left boundary. In order to simplify calculations and interpretation of the results, we simulate potential barriers for electrons with the mass $m$  by three delta functions $U(x) = ({{\hbar ^2 } \mathord{\left/
 {\vphantom {{\hbar ^2 } {2m}}} \right.
 \kern-\nulldelimiterspace} {2m}})\sum\nolimits_{n = 0}^2 {\Omega \delta (x - x_n )} $
  of the same power  $\Omega$  at a distance $d$  from each other at $x_0=0$, $x_1=d$, $x_2=2d$. These points on the $x$  axis demarcate the four regions shown in Fig.\ref{FIG:Fig1}. Delta barrier can be used to model real rather narrow and high potential barrier, while fair estimate  $\Omega  \approx 2mU_b d_b /\hbar ^2 $, where $U_b$  is the height of the barrier and $d_b$   is its width. The electron energy is counted from the vacuum level  $U(x) = U = 0$, which is the same to the left ($x < x_0  = 0$) and to the right ($x > 2d$) of the heterostructure; the effective flat bottom of the heterostructure wells ($0 < x < 2d$) is located at the potential energy  $U(x) = \tilde U < 0$. 

The basis wave functions  $\psi (E,x)$ of the one-dimensional stationary problem of scattering of a wave with the energy $E$  incident from the left are solutions of the Schrodinger equation of the system under consideration and are given by the expressions
\begin{equation}\label{eq:math:1}
\psi (E,x) = \left\{ {\begin{array}{*{20}c}
   {A_{0E} \operatorname{e} ^{ikx}  + B_{0E} \operatorname{e} ^{ - ikx} ,\quad \quad \quad \quad x < 0,}  \\
  \!\!\!{A_{nE} e^{i\tilde k(x - x_{n'} )}\!  +\! B_{nE} \operatorname{e} ^{ - i\tilde k(x - x_{n'} )} ,\;x_{n'}\!  \leqslant \!x\! \leqslant\! x_n,}  \\
   {A_{3E} \operatorname{e} ^{ik(x - x_2 )} ,\quad \quad \quad \quad \quad \quad \quad x > x_2 .}  \\
 \end{array} } \right.
\end{equation}
where $n=1,2$; $n'=n-1$, $k = \hbar ^{ - 1} \sqrt {2mE}$
  is the wave number outside the heterostructure, $\tilde k = \hbar ^{ - 1} \sqrt {2m(E - \tilde U)}$
  is the wave number inside the potential wells, $A_{jE}$  and $B_{jE}$  are the partial amplitudes of plane monochromatic waves propagating, respectively, to the right and left in the regions  $j=0,1,2,3$, and $B_{3E}=0$  (the wave arriving from the right is absent), 
  $A_{0E}  = \hbar ^{ - 1} \sqrt {m/2\pi k}$
  (which provides normalization of $\psi (E,x)$  to the energy  $\delta$-function).
  
The transfer matrix method \cite{Peis2008A}-\cite{Peis2008B} allows one to connect seven partial amplitudes of four regions by linear relations
\begin{equation}\label{eq:math:2}
\left( {\begin{array}{*{20}c}
   {A_{n + 1E} }  \\
   {B_{n + 1E} }  \\

 \end{array} } \right) = M_{nef} \left( {\begin{array}{*{20}c}
   {A_{0E} }  \\
   {B_{0E} }  \\

 \end{array} } \right),
\end{equation}
where  $M_{nef}  = L^{ - 1} M_\Omega  M^n L$, $n=0,1,2$, $M = M_\Omega  M(d)$,
\[
L = \left( {\begin{array}{*{20}c}
   1 & 1  \\
   {ik} & { - ik}  \\

 \end{array} } \right),\quad M_\Omega   = \left( {\begin{array}{*{20}c}
   1 & 0  \\
   \Omega  & 1  \\

 \end{array} } \right),
\]
\[
M(d) = \left( {\begin{array}{*{20}c}
   {\cos \tilde kd} & {\tilde k^{ - 1} \sin {\kern 1pt} \,\tilde kd}  \\
   { - \tilde k\sin \tilde kd} & {\cos \,\tilde kd}  \\

 \end{array} } \right)
\]
and express all partial amplitudes in terms of the amplitude of the incident wave  $A_{0E}$. 
In particular, from \eqref{eq:math:2} at $n=2$  we obtain expressions for the amplitudes of the reflected $B_{0E}  = rA_{0E} $
  and transmitted $A_{3E}  = tA_{0E} $
  waves, where
\begin{equation}\label{eq:math:3}
r =  - \frac{{\tilde M_{21} }}
{{\tilde M_{22} }},\quad t = \frac{{\det \tilde M}}
{{\tilde M_{22} }}
\end{equation}
 $r$ - reflection amplitude, $t$  - transmission amplitude,  $\tilde M_{il} $ - matrix elements of a two-dimensional $(i,l = 1,2)$
  effective transfer matrix  $\tilde M \equiv M_{2ef}  = L^{ - 1} M_\Omega  M^2 L$.
  
\begin{figure}[h]
\includegraphics[width=8.0 cm]{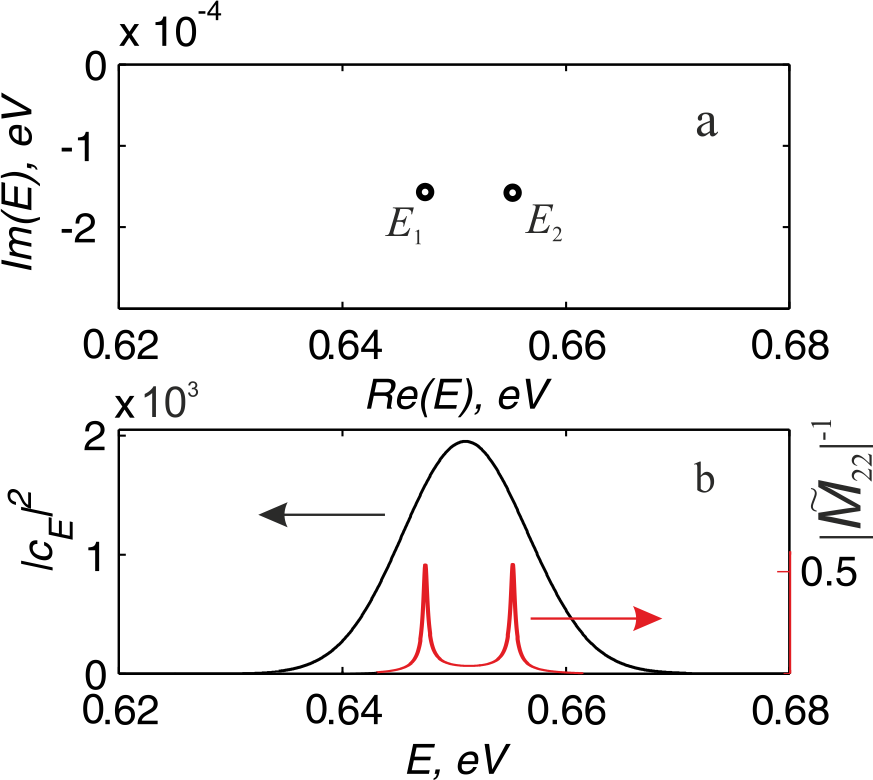}\\
  \caption{(Color online) a) the zeros of $\tilde M_{22}  = 0$  lie in the lower half-plane of the complex energy and for the doublet  lowest above the vacuum level of the heterostructure in Fig.~\ref{FIG:Fig1} (at  $d=125$ \AA,  $\tilde U=-4$ eV, $\Omega=18.9$ \AA$^{-1}=10$  a.u.) are equal  $E_1  = (0.647 - i1.576 \cdot 10^{ - 4} )$
eV and $E_2  = (0.655 - i1.576 \cdot 10^{ - 4} )$
  eV (i.e., the lifetimes of quasi-stationary states $\tau _1  = \hbar /ImE_1  = {\text{4}}{\text{.175}} \cdot 10^{ - 12} $
  s and $\tau _2  = \hbar /ImE_2  = {\text{4}}{\text{.175}} \cdot 10^{ - 12} $ s; 
  b) the red curve with two peaks and the right scale depict the dependence $|\tilde M_{22} |^{ - 1} $
  on the real energy $E$  for this system, the maxima $|\tilde M_{22} |^{ - 1} $
  are at $E_1  = 0.647$ eV, $E_2  = 0.655$ eV; the black curve with one maximum and the left scale depict the square of the modulus of the spectral function $c_E$  of the incident wave packet at optimal for the heterostructure of Fig.~\ref{FIG:Fig1} and Fig.~\ref{FIG:Fig2}a parameters:  $E_C  = \hbar ^2 k_C^2 /2m=0.651$
 eV, $k_C  = 0.414$ \AA$^{-1}=0.219$ a.u., $x_C  =  - 5000$\AA,  $\Delta x=400$\AA, $C_0  = 3.755 \cdot 10^3 $ m$^{-1/2}= 2.73 \cdot 10^{ - 2} $  a.u., $n(x_C,0)=|\Psi(x_C,0)|^2=C_0^2  = 1.41 \cdot 10^7$
  m$^{-1}$ $=7.46\cdot 10^{ - 4} $ a.u. (expressions \eqref{eq:math:4} and \eqref{eq:math:5}).
   Atomic unit of length 1  a.u. $(x) = 0.529 \cdot 10^{ - 10} $ m, atomic unit of time 1 a.u. $ = 2.419 \cdot 10^{ - 17} $ s, 
   atomic unit of probability density 1a.u. $(n) = 1.89\cdot10^{10}$ m$^{-1}$,  atomic unit of probability current density 1 a.u. $(j)=4.1\cdot10^{16}$ s$^{-1}$.}\label{FIG:Fig2}
\end{figure}

Hence, it can be seen that all partial amplitudes (except for  $A_{0E}$), as well as the amplitudes of reflection and transmission, can have pole singularities, which are determined by the zeros of the matrix element  $\tilde M_{22}  = 0$, that is, they can have a resonance character near quasi-stationary levels. Complex roots of the equation $\tilde M_{22}  = 0$  and quasi-stationary levels are grouped into doublets   $E_p  = E'_p  + iE''_p $ $(p = 1,2)$ (Fig.\ref{FIG:Fig2}a). The real parts of pairs of close roots $E'_1  = \operatorname{Re} E_1 $
  and  $E'_2  = \operatorname{Re} E_2 $
 give the energies of quasi-stationary levels. The imaginary parts of the roots $E_1 ^{\prime \prime }  = ImE_1 $
  and $E_2 ^{\prime \prime }  = ImE_2 $
  determine the spectral widths and lifetimes  $\tau _1  = \hbar /E_1 ^{\prime \prime } $
 and  $\tau _2  = \hbar /E_2 ^{\prime \prime } $
 of these quasi-stationary states. The dependence $|\tilde M_{22} |^{ - 1} $
  on the real energy $E$ in the vicinity of the doublet has two close peaks, the widths of which are of the order of  $E_1 ^{\prime \prime } $
 and $E_2 ^{\prime \prime } $
  (Fig.\ref{FIG:Fig2}b).

Let an electronic Gaussian wave packet fall on the heterostructure from the left, the wave function of which at the initial time $t=0$  has the form
\begin{equation}\label{eq:math:4}
\Psi (x,\;0) = C_0 \exp \left( {ik_C x - \frac{{(x - x_C )^2 }}
{{2(\Delta x)^2 }}} \right),
\end{equation}
where $C_0=1/\sqrt{\Delta x\sqrt{\pi}}$, $x_C<0$  is the initial coordinate of the center of the packet, $\Delta x$  - the initial spatial width of the packet, $C_0$  - the initial amplitude of the packet,  $k_C$ - the wave number of the spectral center of the packet, which corresponds to the energy  $E_C  = \hbar ^2 k_C^2 /2m$, in the absence of a scattering potential, the packet moves with the group velocity  $v_C  =\hbar k_C / m$.

The spectral function of the wave packet \eqref{eq:math:4} is determined from the stationary wave functions $\psi(E,x)$  of the scattering problem
\begin{equation}\label{eq:math:5}
c_E=\int_{-\infty}^{\infty}\psi^*(E,x)\Psi(x,0)dx.
\end{equation}
The parameters of the wave packet \eqref{eq:math:4} are chosen so that at  $t=0$, the packet is located far enough to the left of the heterostructure and that its spectral function $c_E$ also has an almost Gaussian form and overlaps mainly only two considered quasi-stationary levels (Fig.\ref{FIG:Fig2}b). It was shown in \cite{Peis2008A} that this can be easily done by satisfying the conditions
\begin{equation}\label{eq:math:6}
k_C^{-1}\ll\Delta x\ll|x_C|\ll k_C(\Delta x)^2,
\end{equation}
then 
\begin{equation}\label{eq:math:7}
c_E  \approx \left\{ {\begin{array}{*{20}c}
   {\dfrac{1}
{\hbar }\sqrt {\dfrac{{m\Delta x}}
{{\sqrt \pi  k}}} {e} ^{ - \frac{{(\Delta x)^2 }}
{2}(k - k_C )^2 } e^{ix_C \left( {k_C  - k} \right)} ,\;E \geqslant 0}  \\
   {0,\quad \quad \quad \quad\quad\quad \quad \quad \quad \quad \quad \quad E < 0}  \\
 \end{array} } \right.
\end{equation}
and the evolution of the packet is mainly determined by the contribution of the energy region  $E_{\min }  < E < E_{\max } $, which includes the selected doublet, but is far from the neighboring doublets. Therefore, at subsequent times, the nonstationary wave function is given by the integral
\begin{equation}\label{eq:math:8}
\Psi (x,t) = \int\limits_{E_{\min } }^{E_{\max } } {c_E e^{ - iEt/\hbar } \psi (E,x)dE} 
\end{equation}

We are interested in the probability density
\begin{equation}\label{eq:math:9}
n(x,t) = \left| {\Psi (x,t)} \right|^2  = \int {\int {\rho _{EE'} (t)} n_{EE'} (x)dEdE'} 
\end{equation}
and the probability current density
\begin{equation}\label{eq:math:10}
\begin{gathered}
  j(x,t) = \frac{{i\hbar }}
{{2m}}\left( {\Psi (x,t)\frac{{d\Psi ^* (x,t)}}
{{dx}} - \Psi ^* (x,t)\frac{{d\Psi (x,t)}}
{{dx}}} \right) =  \hfill \\
  \quad \quad \quad \quad \quad \quad \quad \quad \quad \quad \quad \quad  = \int {\int {\rho _{EE'} (t)j_{EE'} (x)} dEdE'},  \hfill \\ 
\end{gathered} 
\end{equation}
where 
\begin{equation}\label{eq:math:11}
n_{EE'} (x) = n_{E'E}^ *  (x) = \psi (E,x)\psi ^ *  (E',x)
\end{equation}
\begin{equation}\label{eq:math:12}
j_{EE'} (x) = \frac{{i\hbar }}
{{2m}}\left( {\psi (E,x)\frac{{d\psi ^ *  (E',x)}}
{{dx}} - \psi ^ *  (E',x)\frac{{d\psi (E,x)}}
{{dx}}} \right)
\end{equation}
are "matrix elements" of density $n_{EE'}$
  and current density $j_{EE'} (x)$  \cite{Land1977}, and
\begin{equation}\label{eq:math:13}
\rho _{EE'} (t) = \rho _{E'E}^ *  (t) = c_E c_{E'}^ *  e^{ - i(E - E')t/\hbar } 
\end{equation}
is "density matrix" in a "pure" quantum-mechanical state  $\Psi(x,t)$. 

When quasi-stationary states are excited by electromagnetic radiation with subsequent photoemission, especially from ''inside'' the heterostructure, the states of electrons are ''mixed'' and the elements of the density matrix do not have the form \eqref{eq:math:13}, but must be determined from the corresponding kinetic equations.

The density of the electric charge is $\rho _e (x,t) = en(x,t)$
  and of the electric current is  $j_e (x,t) = ej(x,t)$, where $e$  is the electron charge. Differentiating \eqref{eq:math:9}, \eqref{eq:math:10} and applying the nonstationary Schrodinger equation, it is easy to make sure that the law of conservation of the probability density and charge ${{\partial j(x,t)} \mathord{\left/
 {\vphantom {{\partial j(x,t)} {\partial x = }}} \right.
 \kern-\nulldelimiterspace} {\partial x = }} - {{\partial n(x,t)} \mathord{\left/
 {\vphantom {{\partial n(x,t)} {\partial t}}} \right.
 \kern-\nulldelimiterspace} {\partial t}}$
  is satisfied at every point in space at every moment of time.

\section{OSCILLATIONS AND WAVES OF CHARGE AND CURRENT DENSITIES.    EVALUATION FORMULAS AND CALCULATION RESULTS}

\subsection{ Pole Contribution Estimation}
Our main goal here is to demonstrate and explain the regular space-time oscillations of the quantities $n(x,t)$  and  $j(x,t)$, caused by the population and decay of quasi-stationary states of a double quantum well after scattering of a Gaussian wave packet by this well. Substituting \eqref{eq:math:1} and \eqref{eq:math:4} in \eqref{eq:math:5}, \eqref{eq:math:8} - \eqref{eq:math:10} and performing numerical integration in the range of interest, one can obtain a series of figures (Fig. 3-Fig. 10) illustrating the details of the phenomenon.

These oscillations can be described analytically with sufficient accuracy by estimating the integral \eqref{eq:math:8} using the developed by G.F. Drukarev in 1951 \cite{Druk1951} a variant of the saddle point (the fastest descent) method, that allows one to select and evaluate the contributions of the main poles of the scattering amplitudes in the desired integral value \cite{Baz1969}, as was done in \cite{Peis2008A}-\cite{Peis2008B}.
A brief explanation of the essence of this method and the main formulas for calculating the contributions of the saddle points and poles of the integrands are given in Appendix A.

The result of applying the saddle point method strongly depends on the width and position of the saddle in the complex plane, its distance from the origin, as well as on the form of the spectral function of the packet, scattering amplitudes, and on the location and type of their features. The position on the complex plane of the mentioned poles, branch points and other features of the characteristics of stationary scattering and spectral function does not depend on time and coordinates (see (Fig.16) in Appendix A).  However, the saddle points, and with them the lines of type I, for a fixed $x$  move with time $t$  to the origin, usually according to the law $k_S\sim 1/t$, capturing the singular points of stationary scattering in sectors II or III. This determines the appearance of threshold conditions for $x$  and  $t$, under which the singularities make a noticeable contribution to the integrals, providing the manifestation in the form of an envelope of the wave function $\Psi(x,t)$  of various moving maxima, fronts, etc.

In the case under consideration, the saddle points of the exponents in the integrand \eqref{eq:math:8} are responsible for the formation of thresholds and leading pulses of reflection and transmission of the main body of the scattered wave packet, in principle, their contribution can be estimated using (A2). The pole features of $\psi(E,x)$  (i.e., of amplitudes $A_{jE}$  and  $B_{jE}$) are responsible for the formation of the modulation profile of the functions  $\Psi(x,t)$,  $n(x,t)$ and  $j(x,t)$, which can oscillate and slowly decay in time and space due to the rather slow oscillatory decay of the superposition of quasi-stationary states in a double quantum well, which turned out to be populated after the departure of the main body of the wave packet. Their main contribution is proportional to the sum of the residues (A3) at the poles of the integrands \eqref{eq:math:8}. In the space-time regions of the steady oscillations, far enough beyond the thresholds and leading scattering pulses of the main body of the packet (when the saddle point and straight line I turn out to be to the left of the spectral center of the initial wave packet and the poles of the scattering amplitudes), these pole contributions can be large in comparison with other contributions and the wave function is approximately proportional to superpositions of damped traveling waves
\begin{widetext}
\begin{equation}\label{eq:math:14}
\Psi (x,t)\approx \begin{cases}
&\!\!\!\Psi _0 (x,t)+\sum\nolimits_{p = 1}^2 {\tilde B_{0E_p} e^{ - ik_p x - iE_p t/\hbar }  ,\quad x < 0},  \\
&\!\!\!\Psi _n (x,t)+\sum\nolimits_{p = 1}^2 \left( \tilde A_{nE_p } e^{i\tilde k_p (x-x_{n'}) }  + \tilde B_{nE_p }  e^{ - i \tilde k_p (x-x_{n'}) } \right)e^{- iE_p t/\hbar }, \quad n=1,2,\;\;n'=n-1,\;\;x_{n'} \leq x < x_n,   \\ 
&\!\!\!\Psi _3 (x,t) + \sum\nolimits_{p = 1}^2 {\tilde A_{3E_p } e^{ik_p (x - x_2) - iE_p t/\hbar } } ,\quad x > x_2. 
\end{cases}
\end{equation}
\end{widetext}
The terms  $\Psi_0(x,t)$,  $\Psi_{n}(x,t)$,  $\Psi_3  (x,t)$ come from the contributions of those parts of the integration contour of the fastest descent that are far from the poles of the scattering amplitudes; they are smooth functions of $x$  and  $t$ with relatively small magnitude in the regions under consideration  \cite{Peis2008A}-\cite{Peis2008B}. The coefficients $\tilde A_{nE_p }  = |\tilde A_{nE_p } |e^{i\alpha _{np} }$
  ($n=1,2,3$ ) and $\tilde B_{nE_p }  = |\tilde B_{nE_p } |e^{i\beta _{np} } $  ($n=0,1,2$) are proportional to the residues of the integrand \eqref{eq:math:8} at the poles $E_p  = E'_p  + iE''_p $ ($p=1,2$) of the partial amplitudes $A_{nE}$  and $B_{nE}$  from \eqref{eq:math:1} with taking into account the explicitly written coordinate-time exponents, and the complex wave numbers are equal  $k_p  \equiv k(E_p {\kern 1pt} ) = \hbar ^{ - 1} \sqrt {2mE_p }  = k'_p  + ik''_p $
 and $\tilde k_p  \equiv \tilde k(E_p ) = \hbar ^{ - 1} \sqrt {2m(E_{p}  - \tilde U)}  = \tilde k'_p  + i\tilde k''_p $
 (we choose the root branches so as to satisfy the physical conditions of damping waves in space). We are interested in systems that provide a sufficiently slow damping, for which $E'_p  \gg \left| {E''_p } \right|$
  and  $k'_p  \gg |k''_p |$.
   
Below we present the results of numerical calculations using exact formulas \eqref{eq:math:1} and \eqref{eq:math:8} - \eqref{eq:math:12} the quadratic in $\Psi(x,t)$  values of the probability density $n(x,t) = \left| {\Psi (x,\;t)} \right|^2$
  and current of the probability density $j(x,t)$  of electrons inside and outside the considered heterostructure for the parameters of the wave packet and heterostructure, which are given in the caption to 
  Fig.\ref{FIG:Fig2}. These calculations show that approximation \eqref{eq:math:15} provides a reasonable interpretation and estimation of the considered oscillatory and wave effects.

\subsection{The region inside the double quantum well} 
Inside each of the $n=1,2$  wells of heterostructure at  $x_{n-1}\leq x\leq x_n$, substituting the expressions of the second line \eqref{eq:math:1} into the exact formulas \eqref{eq:math:8} - \eqref{eq:math:11} and performing numerical integration, we obtain figures that demonstrate the probability of finding an electron inside a double quantum well (Fig.\ref{FIG:Fig4} and Fig.\ref{FIG:Fig5}) , quasiperiodic flow between the wells of the wave function and the probability  density (Fig.\ref{FIG:Fig4} and Fig.\ref{FIG:Fig6}), as well as the corresponding behavior of the probability current density (Fig.\ref{FIG:Fig7}).

The probability of finding an electron inside a double well $P(t) = \int_0^{2d} {| {\Psi (x,\;t)} |^2 dx} $
  with time first increases rather quickly and then decreases relatively slowly according to a law close to exponential, while  similar probabilities of finding an electron inside each of the two wells $P_n (t) = \int_{(n-1)d}^{nd} {| {\Psi (x,\;t)}|^2 dx} $
  oscillate with the difference frequency of the doublet  $\omega  \equiv \omega _{12}  = (E'_2  - E'_1 )/\hbar $
 and with a period $T = 2\pi /\omega _{12}  = 2\pi \hbar / (E'_2  - E'_1 ) \approx 5.27 \cdot 10^{ - 13}$  s$\approx2.18 \cdot 10^{ 4}$ a.e.$\approx22000$ a.e.
  almost in antiphase with each other (Fig.\ref{FIG:Fig4}):

\begin{figure}[h]
\includegraphics[width=6.5 cm]{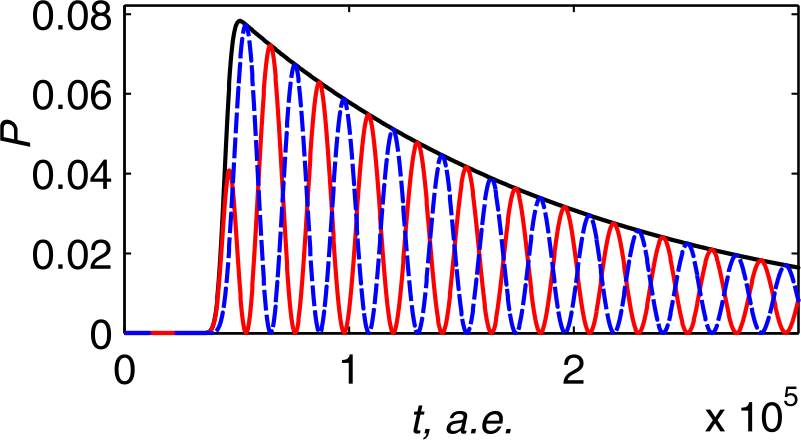}\\
  \caption{(Color online) Time dependence of the probabilities of finding an electron: inside the double well $P(t)$  (black line), in the left well $P_1(t)$  (red line), in the right well $P_2(t)$  (blue line). Time is in atomic units  1 a.u$=2.419 \cdot 10^{ - 17}$ s.}\label{FIG:Fig4}
\end{figure} 

The population of quasi-stationary states occurs approximately during the time of reflection and transmission of the main body of the wave packet, which is equal in order of magnitude to $
\Delta t \sim d/v_C  = md/\hbar k_C  \sim 10^3$ a.e.$\ll \tau$, where  $v_C  = \hbar k_C /m$. Note that the rate of increase in the quantity $P(t)$  (the time $\Delta t$  of penetration of an electron into the well) depends much weaker on the quantity $\Omega$  than the rate of the subsequent decrease (the time $\tau$  of decay of quasi-stationary states). Exponential approximation of the decay part of the curve Fig.\ref{FIG:Fig4} gives the relaxation time of the population of quasi-stationary states in the heterostructure $\tau  = 3.87 \cdot 10^{ - 12} $
 s $= 1.6 \cdot 10^5 $ a.u.  and effective blur $\hbar /\tau  \approx 1.701 \cdot 10^{ - 4} $ eV which is close to  $\operatorname{Im} E_1  = \operatorname{Im} E_2  = 1.576 \cdot 10^{ - 4} $
 eV (see data Fig.\ref{FIG:Fig2}). The area under the curve $P(t)$  and the maximum value of the probability of finding an electron inside the double well $P_{\max}$  change nonmonotonically with increasing value   $\Omega$ due to the nonmonotonic dependence of the transmission coefficients of the  $\delta$-barrier: the value $P_{\max}$   first increases to a certain maximum value at  $\Omega\sim d^{-1}$, and then decreases (Fig.\ref{FIG:Fig5}), but the length of the exponential "tail" $P(t)$  in (Fig.\ref{FIG:Fig4}) monotonically increases. The latter is consistent with the statement proved in our works \cite{Peis2008A}-\cite{Peis2008B} that in a heterostructure formed by  $\delta$-barriers of the same power $\Omega$  located at a distance $d$  from each other, the lifetime $\tau_n$  of the  $n$-th quasi-stationary state increases with increasing $\Omega$  (and $d$), but decreases with increasing $n$  as  $\tau _n  \propto m\Omega ^2 d^4 /(n + 1)^3 $. Hence it follows that for the maximum realization of the studied effects, it is desirable to select the optimal values of all parameters of the problem (see the caption to (Fig.\ref{FIG:Fig2})), so that both $P_{\max}$  and $\tau_n$  are as large as possible.

\begin{figure}[h]
\includegraphics[width=6.5 cm]{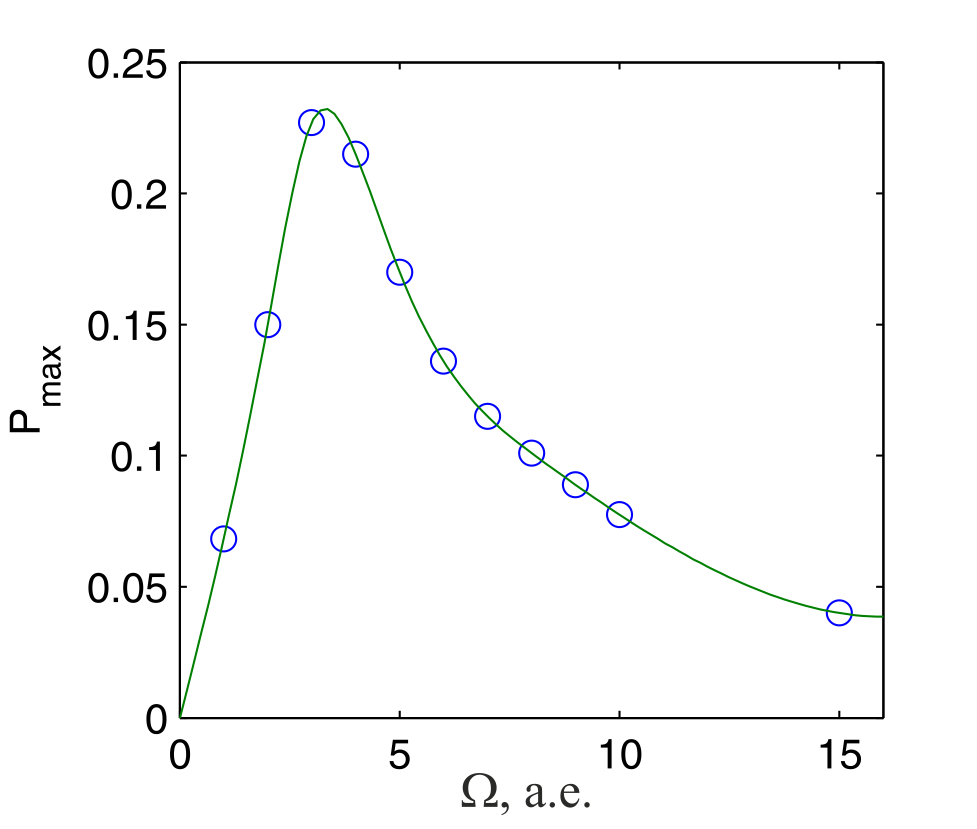}\\
  \caption{(Color online) Maximum probability of finding a particle inside a double well  $P_{\max}$, depending on the barrier power $\Omega$  with other fixed parameters (Fig.\ref{FIG:Fig2}).}\label{FIG:Fig5}
\end{figure}

\begin{figure}[h]
\includegraphics[width=8.5 cm]{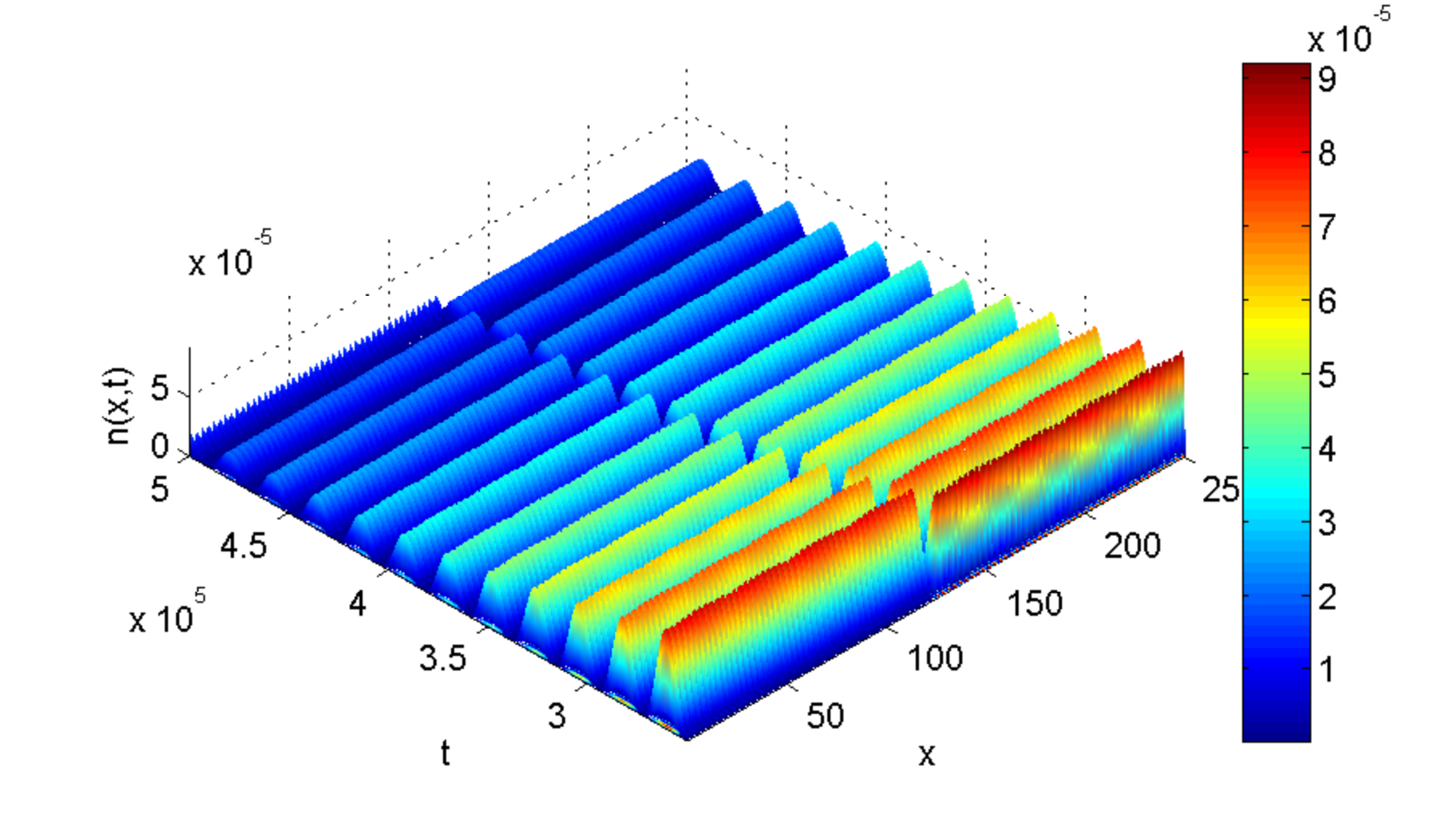}\\
  \caption{(Color online) Coordinate-time dependence of the values (color scale on the right) of the probability density $n(x,t)$  of finding a particle inside the double well  $0\leq x\leq 2d$. Coordinate in angstroms \AA, time and probability density in atomic units.}\label{FIG:Fig6}
\end{figure} 

\begin{figure}[h]
\includegraphics[width=8.5 cm]{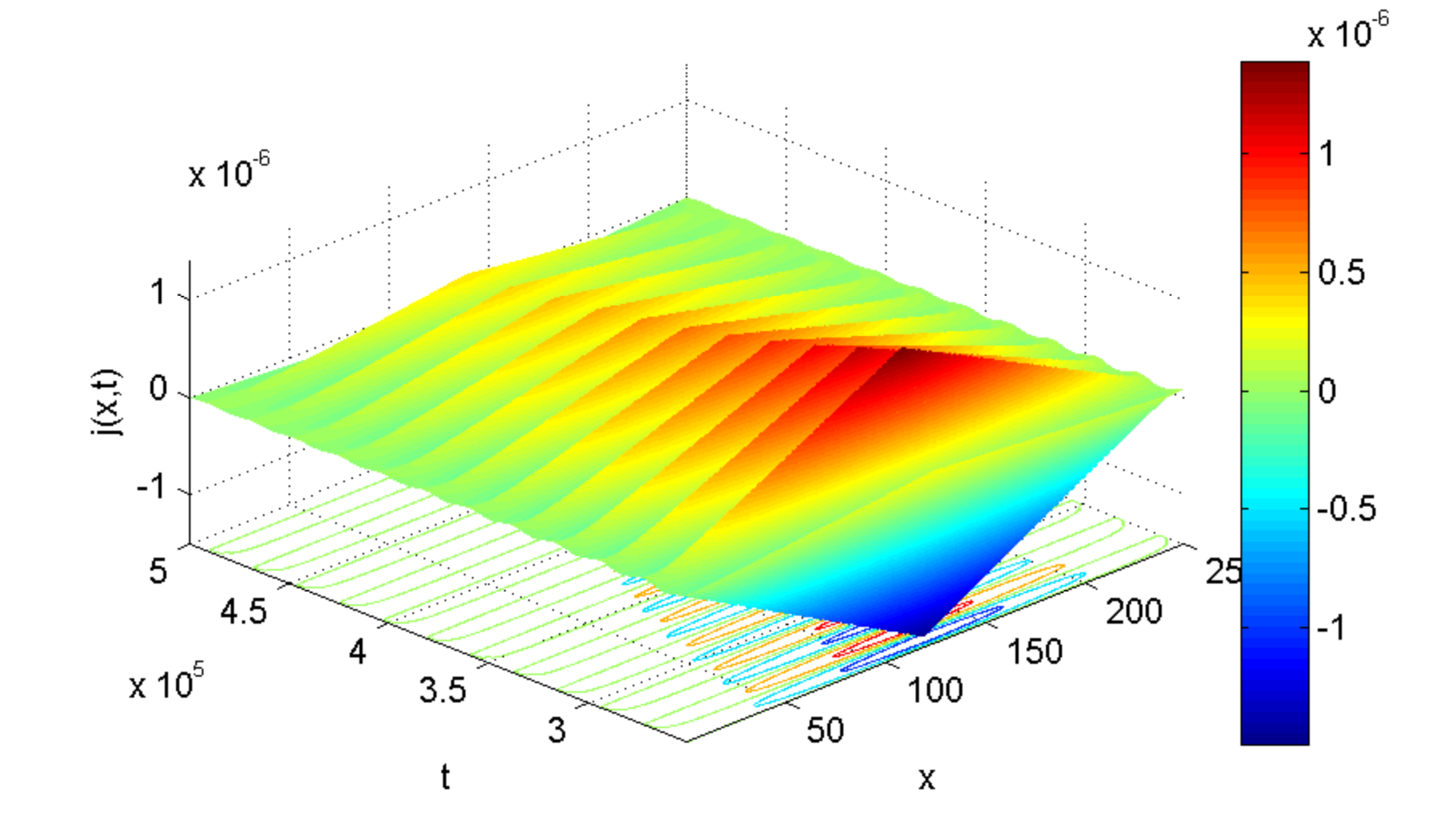}\\
  \caption{(Color online) Coordinate-time dependence of the values (color scale on the right) of the probability current density $j(x,t)$  of a particle inside the double well  $0\leq x\leq 2d$. Coordinate in angstroms \AA, time and probability density in atomic units.}\label{FIG:Fig7}
\end{figure} 

Figures (Fig. 5) and (Fig. 6) are quite well explained by \eqref{eq:math:9}, \eqref{eq:math:10} and the second line \eqref{eq:math:14}, at such values of $t$  and  $x$, at which it is possible to neglect  $\Psi_n(x,t)$.  Complete analytical expressions $n(x,t)$  and $j(x,t)$ in this approximation are given by formulas \eqref{eq:math:B1} and \eqref{eq:math:B2}, which are written out in Appendix B. It can be seen from \eqref{eq:math:B1} and \eqref{eq:math:B2} 
 that, inside the double well, the quantities $n(x,t)$  and $j(x,t)$  undergo spatio-temporal oscillations and weak exponential decay with time. The terms in the first lines of both expressions \eqref{eq:math:B1} and \eqref{eq:math:B2} almost do not change with time  $t$ and coordinate  $x$, the terms in the second lines weakly decay with time, but quickly change along the coordinate   with spatial periods $\tilde \lambda_p  = \pi /\tilde k'_p  \sim \pi /k_C$, which are small in comparison with the width of the wells  $d$. The last four lines in both expressions \eqref{eq:math:B1} and \eqref{eq:math:B2}  describe plane waves traveling to the right and left inside the wells, the corresponding wave-like temporal oscillations $n(x,t)$  and $j(x,t)$  occur with the difference frequency of the doublet  $\omega  \equiv \omega _{12}  = (E'_2  - E'_1 )/\hbar $. In this case, the terms in the third and fourth lines describe traveling waves, the wavelength of which $\tilde \lambda _ -   = 2\pi | {\tilde k'_1  - \tilde k'_2 }|^{ - 1} $ is large in comparison with the width of the wells  $d$; therefore, such terms inside the wells are almost independent of $x$  at a fixed  $t$, the phase velocity of these waves is $\tilde v_ -   =\omega \tilde \lambda _ -  / 2\pi  = (E'_2  - E'_1 )/\hbar | {\tilde k'_1  - \tilde k'_2 } | = 7.189 \cdot 10^5 $ m/s. However, the fifth and sixth lines describe short-wavelength waves traveling towards each other, the wavelength of which $\tilde \lambda _ +   = 2\pi \left| {\tilde k'_1  + \tilde k'_2 } \right|^{ - 1}  \sim \pi /k_C $
  is small compared to the width $d$  of the wells, and the phase velocity of such waves 
 $\tilde v_ +   =\omega \tilde \lambda _ + /2\pi  = (E'_2  - E'_1 )/\hbar | \tilde k'_1  + \tilde k'_2 | \approx 2.202 \cdot 10^3 $ m/s, is small compared to  $\tilde v_-$. 
    Note also that in expression \eqref{eq:math:B1} all terms have almost the same order of magnitude, therefore, in the figure (Fig.\ref{FIG:Fig6}), the coordinate dependence $n(x,t)$  inside the wells is dominated by short-wavelength components with a wavelength $\tilde \lambda _ +   \sim \pi /k_C$, which rather abruptly change their amplitude between the wells. On the contrary, in expression \eqref{eq:math:B2} such short-wave components make a relatively small contribution to the coordinate dependence of $j(x,t)$  in comparison with long-wave components $\tilde \lambda _ - $: the fifth and sixth lines of expression \eqref{eq:math:B2} contain a small factor $| {\tilde k'_1  - \tilde k'_2 } | \ll \,\,k_C $, and the third and fourth lines of expression \eqref{eq:math:B2} contain a large factor $| {\tilde k'_1  + \tilde k'_2 }| \approx 2\,k_C $, therefore, on Figure (Fig. \ref{FIG:Fig7}) the coordinate dependence of $j(x,t)$  inside the wells is very smooth with a break at the boundaries of the wells.

\subsection{ The region outside the double quantum well on the left}

\begin{figure}[h]
\includegraphics[width=8.5 cm]{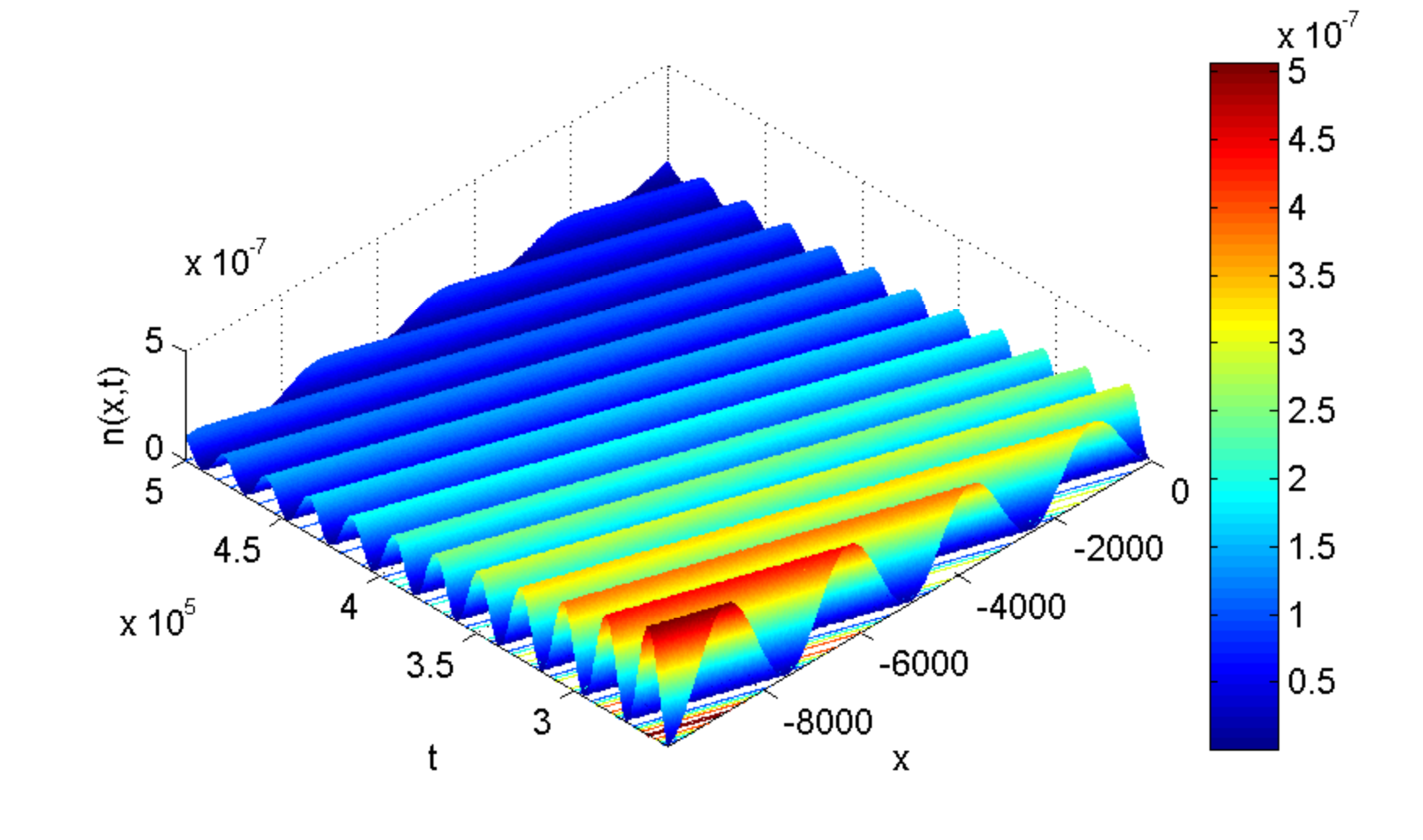}\\
  \caption{(Color online) Wave coordinate-time dependence of the values (color scale on the right) of the probability density $n([,t)$  of finding a particle in the corresponding points of the left half-space  $x<0$. Coordinate  $x$ in angstroms \AA, time and probability density in atomic units.}\label{FIG:Fig8}
\end{figure}

\begin{figure}[h]
\includegraphics[width=8.5 cm]{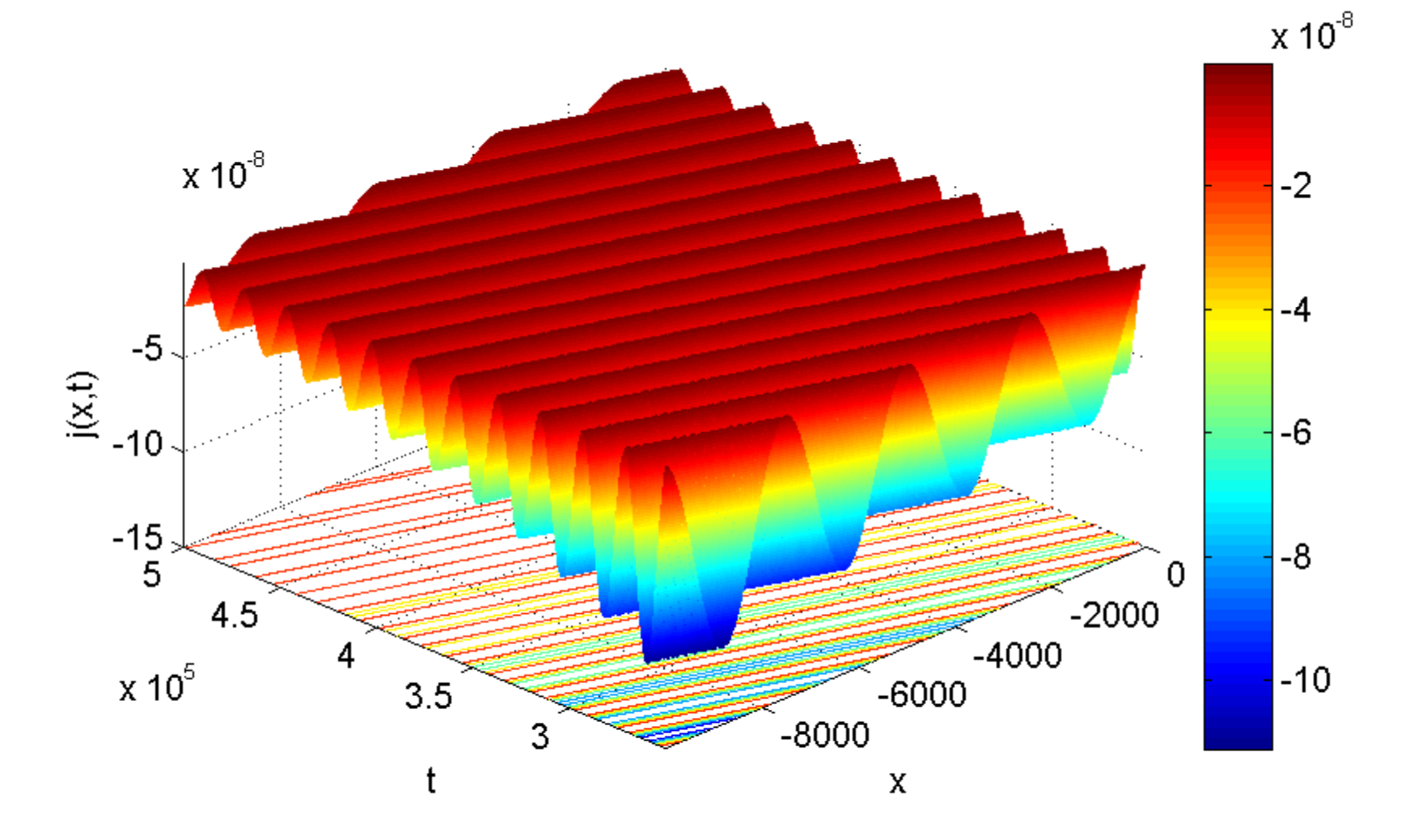}\\
  \caption{(Color online) Wave coordinate-time dependence of the values (color scale on the right) of the current probability density $j(x,t)$   of finding a particle in the corresponding points of the left half-space  $x<0$. Coordinate $x$  in angstroms \AA, time and probability density in atomic units.}\label{FIG:Fig9}
\end{figure}

Similarly, to the left of the double well at  $x<0$, substituting the expression of the first line \eqref{eq:math:1} into the exact formulas \eqref{eq:math:8}-\eqref{eq:math:12} and performing numerical integration, we obtain figures that demonstrate the decaying probability density waves (Fig.\ref{FIG:Fig8}) and current density waves traveling to the left  (Fig.\ref{FIG:Fig9}).

These figures (Fig.\ref{FIG:Fig8}) and (Fig.\ref{FIG:Fig9}) are also quite well explained by \eqref{eq:math:9}, \eqref{eq:math:10} and the first line \eqref{eq:math:15}, at such values of $t$  and  $x$, at which it is possible to neglect  $\Psi_0(x,t)$, that gives the main pole contributions to $n(x,t)$ and $j(x,t)$   in the form of analytical formulas \eqref{eq:math:B3} and \eqref{eq:math:B4} given in Appendix B, which describe the probability density and probability current waves traveling to the left. 
Figures (Fig.\ref{FIG:Fig8}) and (Fig.\ref{FIG:Fig9}) show that to the left of the double well, the wave part of $j(x,t)$  changes almost in antiphase to the wave part of  $n(x,t)$. This is explained by the minus sign in \eqref{eq:math:B4} and the fact that we have $k'_1  \approx k'_2  \approx k_C  = 0.219$ a.u.

\subsection{The region outside the double quantum well on the right}

In the same way, to the right of the double well at  $x>x_2$, after substituting the expression of the third line \eqref{eq:math:1} into the exact formulas \eqref{eq:math:8} - \eqref{eq:math:12}  and numerical integration, figures are obtained that demonstrate the decaying waves of the probability density (Fig.\ref{FIG:Fig9}) and the probability current density traveling to the right.  

\begin{figure}[h]
\includegraphics[width=8.5 cm]{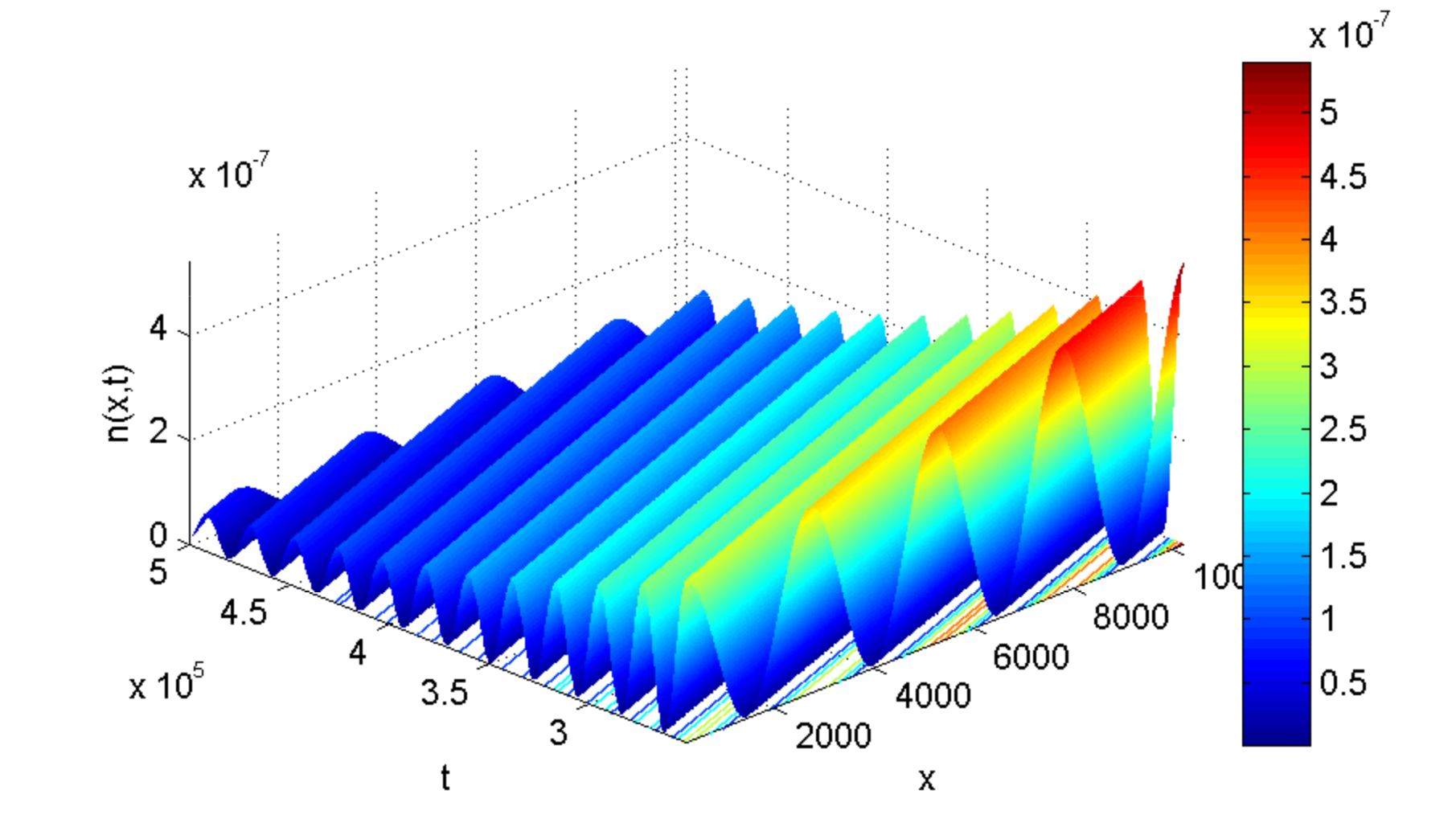}\\
  \caption{(Color online) Wave coordinate-time dependence of the values (color scale on the right) of the probability density $n(x,t)$  of finding a particle in the corresponding points of the right half-space $x>x_2$. Coordinate $x$   in angstroms \AA, time and probability density in atomic units.}\label{FIG:Fig10}
\end{figure}

For the coordinate-time wave dependence of the current density $j(x,t)$  of a particle at the points of the right half-space  $x>x_2$, the figure qualitatively looks like Fig.\ref{FIG:Fig10}, that is, to the right of the double well, the wave parts of $n(x,t)$  and $j(x,t)$  change almost in phase, but in atomic units $j(x,t)$  is less than $n(x,t)$  about a decimal order of magnitude. 

These dependences are also reasonably well explained by \eqref{eq:math:9}, \eqref{eq:math:10}  and the third line \eqref{eq:math:14} at such values of $t$  and $x$  for which it is possible to neglect  $\Psi_3(x,t)$, that gives the main pole contributions to $n(x,t)$ and $j(x,t)$   in the form of analytical formulas \eqref{eq:math:B5} and \eqref{eq:math:B6} given in Appendix B, which describe the probability density and probability current waves traveling to the right. 
 
The noted similarity and difference in behavior of $j(x,t)$  and $n(x,t)$   is explained by the presence in \eqref{eq:math:B6} in comparison with \eqref{eq:math:B5} of factors containing $k'_1  \approx k'_2  \approx k_C  = 0.219$ a.u.

\subsection{The complete picture of generation of probability density and current waves} 

Thus, inside a heterostructure in the form of a double quantum well, oscillations of the electron density and current with the difference frequency of the doublet   $\omega  \equiv \omega _{12}  = (E'_2  - E'_1 )/\hbar $ can occur, which looks like a periodic overflow of the electron wave function $\Psi(x,t)$  and the probability density $n(x,t)$  between the wells (in time almost in antiphase to the left and to the right), so that outside the heterostructure the probability density waves and currents density waves outgoing to the left and to the right are formed. 
 In this case, outside the heterostructure, quadratic in magnitude $\Psi(x,t)$  values  $n(x,t)$ and  $j(x,t)$ oscillate in time with the difference frequency of the doublet  $\omega  \equiv \omega _{12} $, and in space with a wavenumber equal to the difference  $k_{12}  = k'_2  - k'_1$, slowly decaying with decrements determined by the imaginary parts of $E_p$   and  $k_p$.

\begin{widetext}

\begin{figure}[ht]
\includegraphics[width=16 cm]{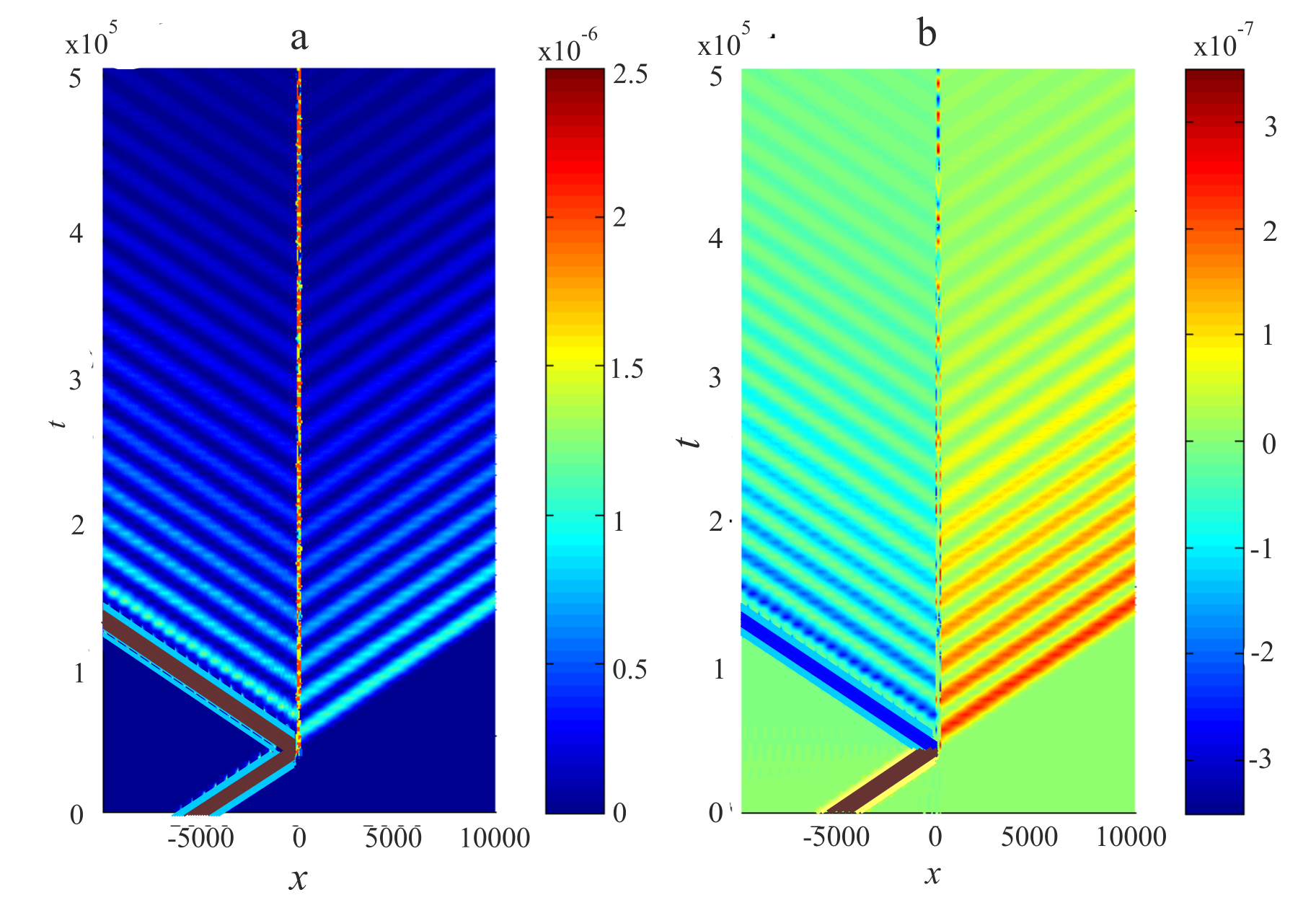}\\
  \caption{(Color online) The calculated relief levels of the quantities a) probability density  $n(x,t)$, b) probability current density $j(x,t)$  in accordance with the color scales to the right of the figures. On this scale, the region inside the double well is not allowed, and the two lower stripes on the left qualitatively represent the main bodies of the incident and reflected wave packets having $n\sim10^{-3}$  a.u. and $j\sim10^{-5}$ a.u. Coordinate $x$  in angstroms \AA, time, probability density and probability current density in atomic units.}\label{FIG:Fig11}
\end{figure}

\end{widetext}

Waves of  $n(x,t)$ and $j(x,t)$  move to the left and to the right with the same velocities  $\operatorname{v}  \approx \lambda /T =  4.79 \cdot 10^5 $ m/s, where the wavelength is $
\lambda  = 2\pi /k_{12}  = 2\pi /|k'_2  - k'_1 | \approx 2480$\AA, and the period of the waves is  $T=2\pi/\omega_{12}=2\pi \hbar/(E'_2-E'_1)\approx2.18\cdot10^4$ a.u. $\approx5.27\cdot10^{-13}$s. The generation of these waves can be represented on Fig.\ref{FIG:Fig11} by level lines on the  $t-x$-plane.

%%%%%%%%%%%%%%%%%%%%%%%
\section{PROLONGATION AND AMPLIFICATION OF WAVE GENERATION}
\begin{figure}[h]
\includegraphics[width=8.5 cm]{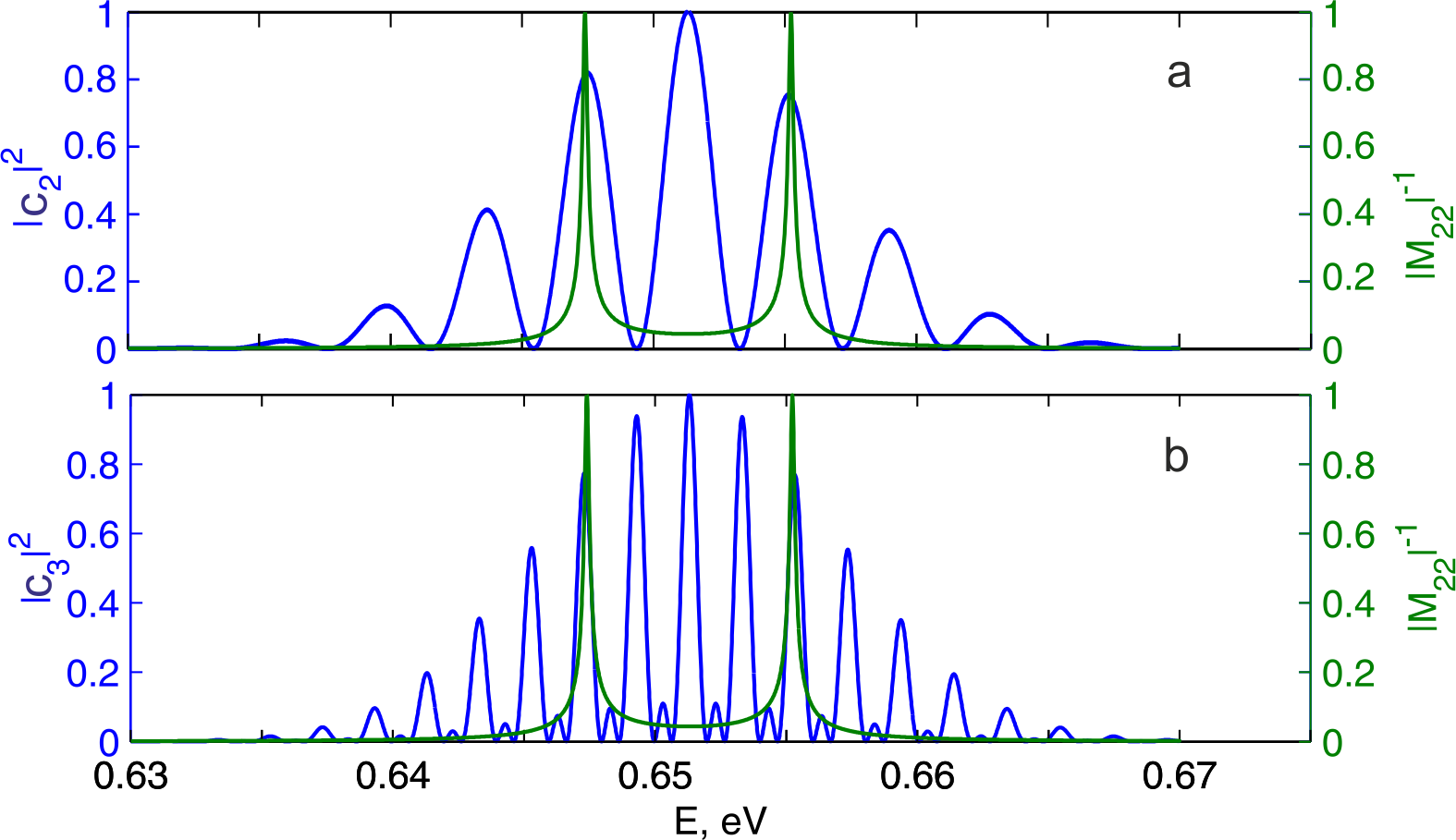}\\
  \caption{(Color online) Spectral functions $c_N(E)$  at $s=2$  favorable for maximizing wave amplification versus resonance peaks  $|\tilde M_{22}|^{-1}$ (cf. (Fig.\ref{FIG:Fig2}b)): a) for $N=2$  resonance are given by the first main maxima of  $|c_2|^2$, b) for   resonance are given by the second main maxima of  $|c_3|^2$. The curves are brought to the same unit scale for ease of comparison}\label{FIG:Fig12}
\end{figure}

In the system under consideration, it is possible to organize a mode of repetition or amplification of the process of radiation of electron probability density and probability current density waves. If we know the space-time periods of the waves under study, then in order to prolong the radiation process and increase the amplitude of the density and current waves, we can form a quasiperiodic sequence of wave packets similar to the original packet \eqref{eq:math:4} to the left of the double quantum well and send them in such a way as to provide an additional resonant pumping of the population of quasi-stationary states of the heterostructure. To prepare such a coherent chain of pulses, one can, for example, use two methods:

1) The first of these methods consists in aligning along the axis $x$ of an equidistant sequence of identical pulses with a spatial period close to a value that is a multiple of the doubled resonant difference wavelength  $\lambda=2\pi/|k_{12}|$. At the initial moment of time  $t=0$, the wave function should be prepared in the form of a spatial sequence of $N$  identical wave packets following the head packet \eqref{eq:math:4}, in which the initial coordinates of the centers  $x_{Cn}=x_C-n\delta x$ are shifted relative to $x_C$  ($\delta x$  is shift period; $n=1,2,...,N$). If these packets almost do not overlap and for each of them conditions \eqref{eq:math:6} and \eqref{eq:math:7} are fulfilled with replacement  $x_C\to x_{Cn}$, then instead of the spectral function $c_E$  in the integrand \eqref{eq:math:8} there appears the spectral function $c_N(E)$  of the entire sequence of packets, which in this case is given by the sum
\begin{equation}\label{eq:math:15}
c_N (E) = c_E \sum\limits_{n = 0}^{N - 1} {{\text{e}}^{ - in\delta x\delta k} }  = c_E {\text{e}}^{ - i(N - 1)\delta x\delta k/2} y_N \left( {\frac{{\delta x\delta k}}
{2}} \right)
\end{equation}
where $\delta k = k_C  - k$, and an interference function
\begin{equation}\label{eq:math:16}
y_N (z) \equiv \frac{{\sin (Nz)}}
{{\sin z}}
\end{equation}
is periodic in $z$  with a period $2\pi$  and has the main extrema  $y_{N\max}=N$ at the values of the argument  $z_{\max}=s\pi$, where $s$  is an integer. In the theory of diffraction gratings, it describes an increase in the amplitude of the resultant wave at its main resonance maxima by $N$  times and its intensity by  $N^2$ times. In expression \eqref{eq:math:16} we have $z(k)\equiv\delta x\delta k/2$  and it is obvious that at the main extrema of the function $y_N(z)$  all exponentials are equal to one under the sum sign, and the entire sum is equal  $N$. In our case, the integration of \eqref{eq:math:8} with  $c_N(E)$ instead of $c_E$  provides a significant contribution of the poles  $k_p  = k'_p  + ik''_p $  of the scattering amplitudes, as for one wave packet, therefore, due to the superposition of $N$  resonant diffracted waves, the function $y_N(z)$  can also provide up to a close to  $N^2$-fold (on conditions $\left| {k''_p } \right| \ll k'_p$) amplification of the wave amplitudes  $n(x,t)=|\Psi(x,t)|^2$ and $j(x,t)$   in comparison with their values for one ($N=1$) wave packet in the corresponding intervals of $x$  and  $t$. This takes place if the points  $k_m$ of the main extrema of the function  $y_N(z(k))$ are close to the points $k_1\approx k'_1$  and $k_2\approx k'_2$  of  the resonance maxima of the moduli of the amplitudes of stationary scattering on the double well, which can be ensured by selecting the value  $\delta x$. Indeed, the period of  $y_N(z(k))$ by argument  $k=k_C-\delta k$ is equal to  $4\pi/\delta x$, when $k$  is counted from  $k_C$, and since our spectral center $k_C$  of the original wave packet is located almost in the middle between the resonance wave numbers $k_1\approx k'_1$  and  $k_2\approx k'_2$, then at the main extrema there should be  $|\delta k| = | k_C  - k_m  | = | k_{12}| /2=\pi/\lambda$, so favorable for maximum amplification values of the shift periods should be close to $\delta x\approx 4\pi s/k_{12}=2s\lambda$ (Fig.\ref{FIG:Fig12}). Weaker amplification of waves can occur at such values of $\delta x$  for which  $N > | {y_N (z(k_1 ))} | \approx | {y_N (z(k_2 ))} | \geq 1$, and the weakening of the sum wave will occur at $| {y_N (z(k_1 ))} | \approx | {y_N (z(k_2 ))} | < 1$. 

\begin{figure}[h]
\includegraphics[width=7.50 cm]{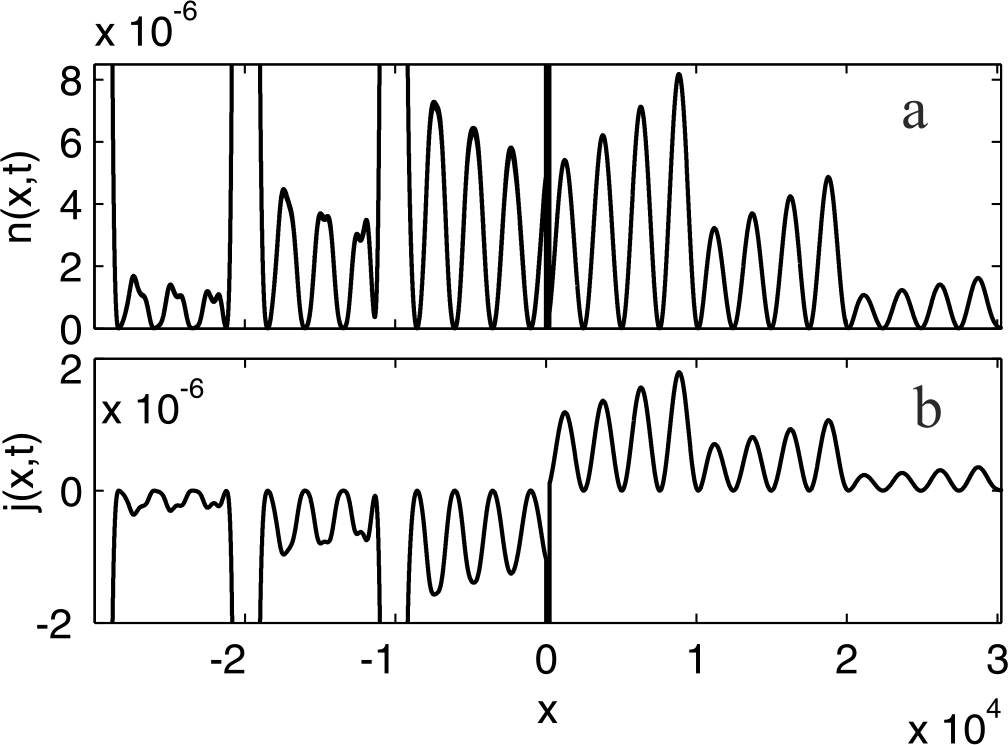}\\
  \caption{ The spatial profile of the resonant amplification of density and current probability waves by a sequence of three ($N=3$) identical wave packets shifted relative to each other by a distance of $\delta x\, \approx 4\pi s/k_{12}  = 2s\lambda $ = 9828 \AA,  $s=2$ at the moment of time $t = 3 \cdot 10^5 $ a.u. $ = 7.26 \cdot 10^{ - 12} $ s (the main bodies of the reflected packets are cut off because they are not of interest to us, they are about an order of magnitude larger than the vertical size of the panels). Coordinate in angstroms \AA, time, probability density and probability current density in atomic units.}\label{FIG:Fig13}
\end{figure} 

To find the period  $\delta x$, we also used a more general method, which is also valid in cases where conditions \eqref{eq:math:6} and \eqref{eq:math:7}  are violated for all sequential packets. Namely, the period $\delta x$  was determined numerically from the points of intersection on the  $E-\delta x$-plane of straight lines  $E=E_1$,  $E=E_2$ with the lines of the main extrema of the spectral function of the entire sequence of wave packets, parametrically depending on  $\delta x$, and calculated not according to \eqref{eq:math:16}, but according to the general formula \eqref{eq:math:5}, in which $\Psi(x,0)$  it is taken equal to the initial the wave function of the entire sequence of wave packets.

\begin{figure}[h]
\includegraphics[width=7.5 cm]{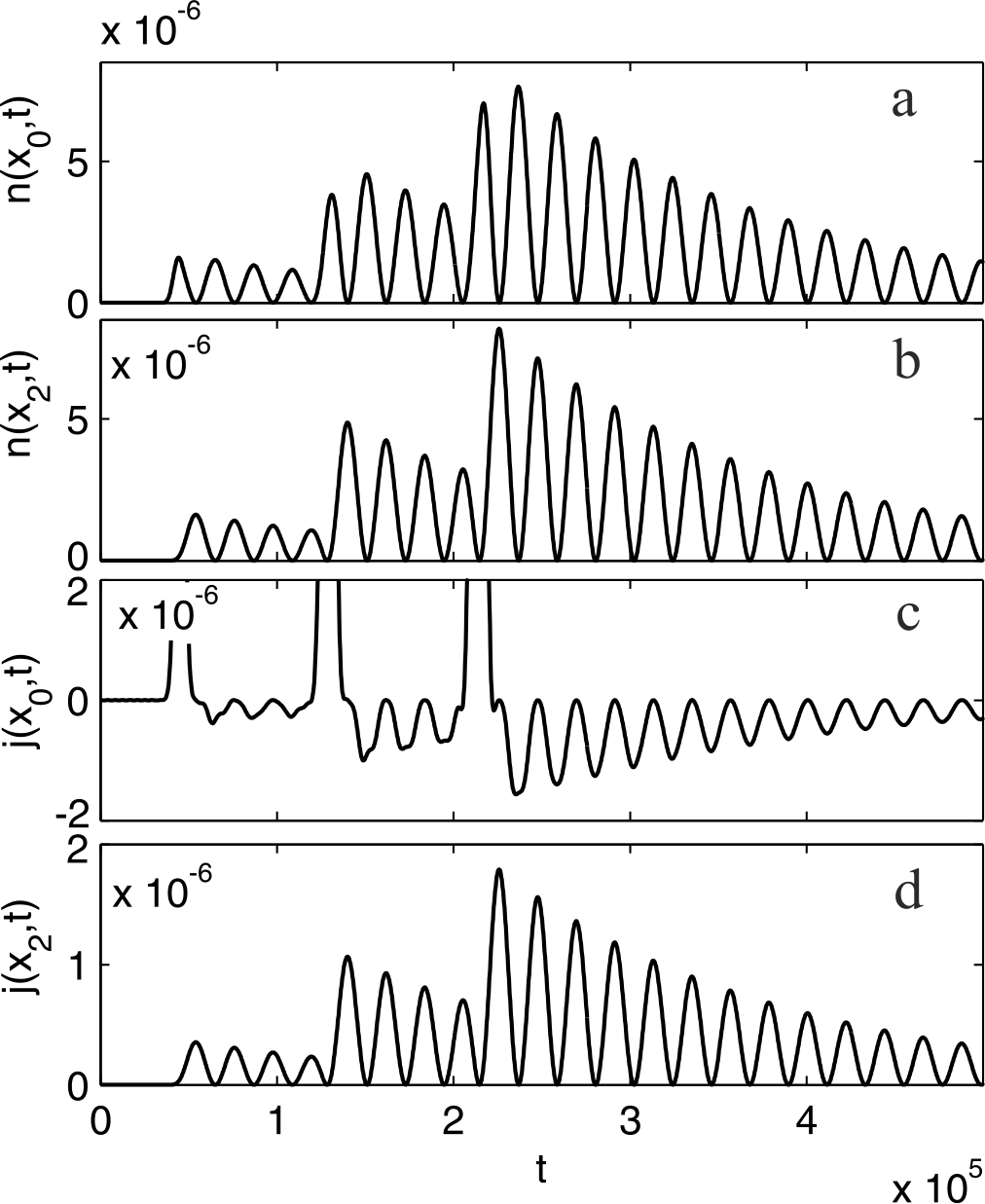}\\
  \caption{ The time profile of the resonant amplification of the probability density and current waves by a sequence of the same three ($N=3$) identical wave packets (at  $s=2$), as in Fig.\ref{FIG:Fig13} using the example of points $x=x_0=0$  on the left  and $x=x_2=2d$  on the  right boundaries of the double quantum well. The time, probability density and probability current density in atomic units.}\label{FIG:Fig14}
\end{figure}

Figures (Fig.\ref{FIG:Fig13}) and (Fig.\ref{FIG:Fig14}) demonstrate the resonant coherent amplification of the probability density and probability current waves by a spatial sequence of three ($N=3$) identical wave packets shifted in space by $\delta x\, \approx 4\pi s/k_{12}  = 2s\lambda $  at  $s=2$.

2) The second method involves the creation of almost identical pulse wave packets in one place sequentially in time with a period close to a multiple of the resonant difference time period $T=2\pi/\omega_{12}$  of the wave using the appropriate time aperture function. In this case, the source of particles should be arranged in such a way that coherent wave impulses of the form \eqref{eq:math:9}, \eqref{eq:math:10}, which follow each other, appear sequentially at the same place with a time period  $\delta t$. If we assume that these packets almost do not overlap and for each of them conditions of the type \eqref{eq:math:6} and \eqref{eq:math:7} are satisfied, then in each time interval  $(N-1)\delta t\leq t\leq N\delta t$ when the $N$  pulses are excited (and also for $t>(N_0-1)\delta t$  if the pulse with the number $N=N_0$  is the last) in the expressions \eqref{eq:math:9} and \eqref{eq:math:10}, instead of the product of spectral functions  $c_E c_{E'}^ *$, a function of the entire sequence of pulses $(c_E c_{E'}^ *)_N$  appears, which is now given by the sum
\begin{equation}\label{eq:math:17}
(c_E c_{E'}^ *  )_N = c_E c_{E'}^ * \sum_{n = 0}^{N - 1} 
e^{i n\delta t(E - E')/\hbar }   =  c_E c_{E'}^ *  e^{i(N - 1)z'} y(z')  
\end{equation}

The interference function $y(z')$  has the same form \eqref{eq:math:17}, but now its argument is equal $z' \equiv z'(E - E') = \delta t(E - E')/2\hbar $, as above the function $y(z')$  is periodic with the period $2\pi$   and has main extrema $|y_{\max}|=N$  at the values of the argument  $z'_{\max }  = s\pi $, where  $s$ is an integer. Integration in \eqref{eq:math:9} and \eqref{eq:math:10} with  $(c_E c_{E'}^ *)_N$ instead of $c_E c_{E'}^ *$  provides the determining contribution of the poles $E_p  = E'_p  + iE''_p $
  of the scattering amplitudes, so that in the corresponding intervals  $x$ and  $t$, in which the $N$  pulses of $n(x,t)$  and $j(x,t)$  are already excited and undergo diffraction, and due to their superposition the function $y(z')$  can now provide only to near  $N$-fold amplification of the wave amplitudes $n(x,t) =| {\Psi (x,\;t)}|^2 $
  and  $j(x,t)$ (on conditions  $|E''_p|\ll E'_p$, otherwise due to attenuation, the amplification is weaker) compared to their values for one ($N=1$) wave packet. This takes place if equalities $E - E' \approx E'_2  - E'_1  = \hbar \omega _{12}  = 2\pi \hbar /T = 2\pi \hbar s/\delta t $   are satisfied at the poles  $E_p  = E'_p  + iE''_p $, which can be ensured by selecting a value $\delta t$  close to a value that is a multiple of the period of these waves $\delta t=\delta T$. Weaker amplification of waves can occur at values of $\delta t$  for which  $N>|y_N(z')|\geq 1$, and there will be attenuation of waves at  values of  $\delta t$ for which  $|y_N(z')|< 1$.
  
The period $\delta t$  favorable for amplification can also be found numerically by $\delta t$   vertically shifting the patterns of Fig.\ref{FIG:Fig11}a) and/or Fig.\ref{FIG:Fig11}b) until parallel oblique lines of maxima (and minima) of the shifted and not shifted patterns are superimposed on each other after the required number  $s$ of periods for the required the number $N$  of wave packets.

In relation to $\Psi(x,t)$  the waves $n(x,t) = | {\Psi (x,\;t)} |^2$
  and $j(x,t)$  are a kind of "intensity waves", in case 2) they experience amplification only by a factor of  $N$, in contrast to the previous case 1) and the situation in the theory of diffraction gratings, which provide an increase in intensity by a factor of  $N^2$.
  
\begin{figure}[h]
\includegraphics[width=7.5 cm]{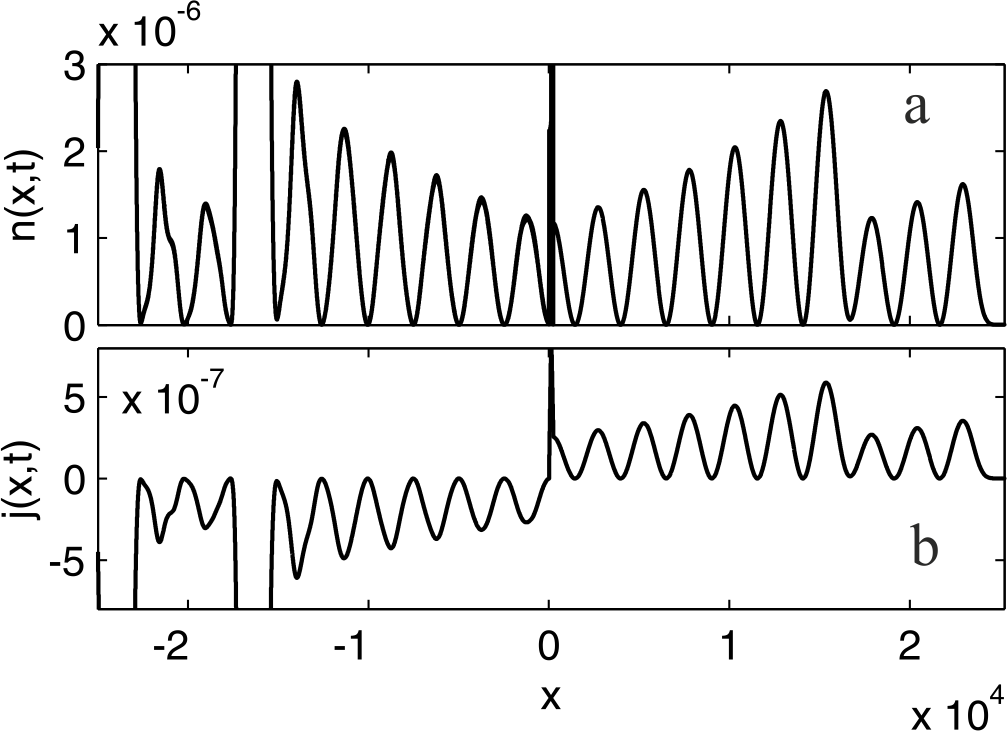}\\
  \caption{ The spatial profile of the resonant amplification of density and current probability waves by a sequence of two ($N=2$) identical wave packets shifted in time relative to each other by $\delta t=sT= 65448$ a.e.,  $s=3$,  at the moment of time $t = 3 \cdot 10^5 $
 a.u. $ = 7.26 \cdot 10^{ - 12} $ s (the main bodies of the reflected packets are cut off because they are not of interest to us, they are about an order of magnitude larger than the vertical size of the panels). Coordinate in angstroms \AA, time, probability density and probability current density in atomic units.}\label{FIG:Fig15}
\end{figure}

\begin{figure}[h]
\includegraphics[width=7.5 cm]{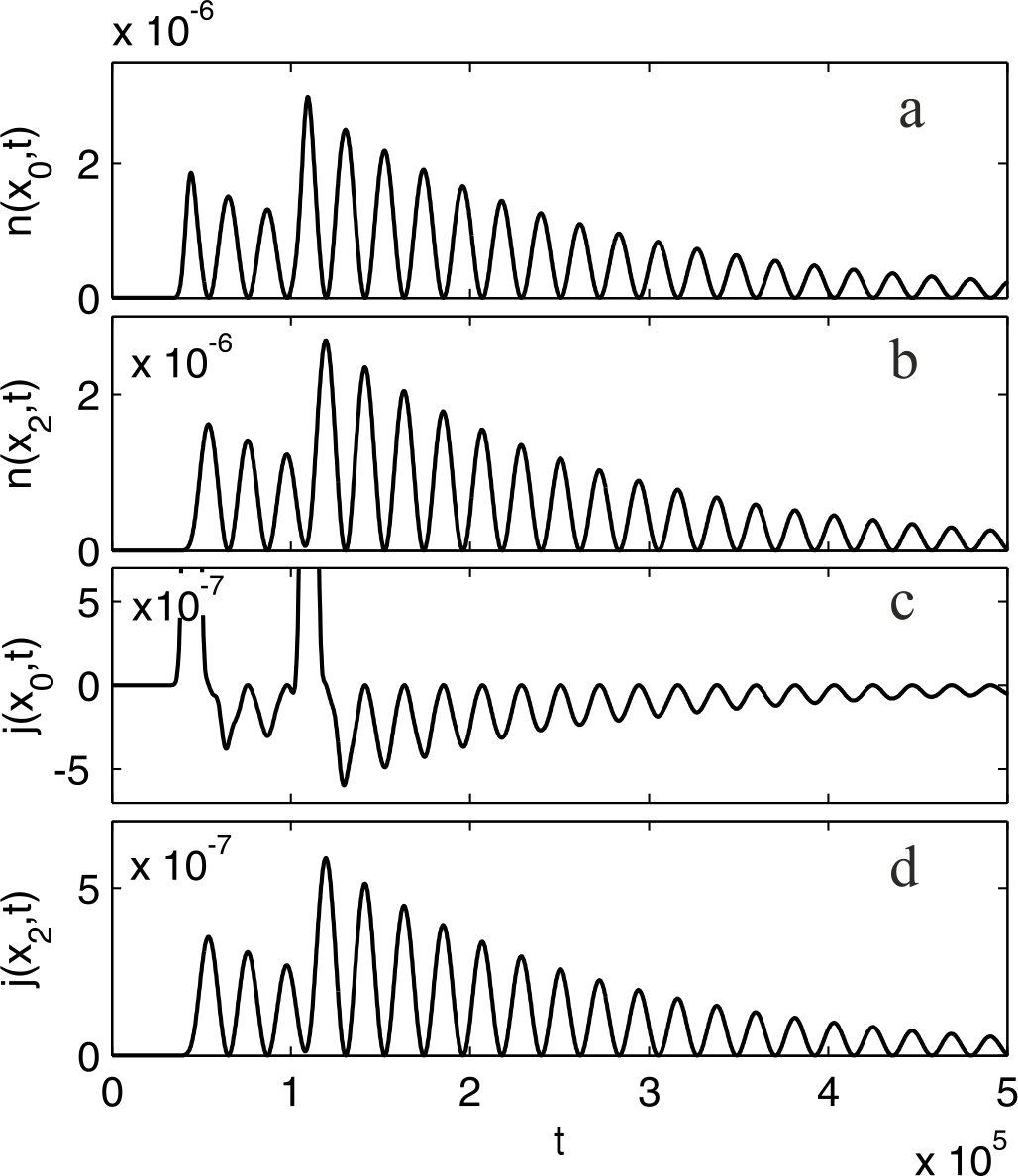}\\
  \caption{ The time profile of the resonant amplification of the probability density and current waves by a sequence of the same two ($N=2$) identical wave packets (at $s=3$), as in Fig.\ref{FIG:Fig15} using the example of points $x=x_0=0$  on the left  and $x=x_2=2d$  on the  right boundaries of the double quantum well. The time, probability density and probability current density in atomic units.}\label{FIG:Fig16}
\end{figure}

Figures (Fig.\ref{FIG:Fig15}) and (Fig.\ref{FIG:Fig16}) demonstrate the resonant coherent amplification of the probability density and current waves by the temporal sequence of two ($N=2$) identical wave packets shifted in time by $\delta t=sT=2\pi s/|\omega_{12}|$  at  $s=3$.

\section{ CONCLUSION}

The generation probability density and probability current density waves of electrons in the range of terahertz frequencies and micrometer wavelengths is of interest from the point of view of various applications of micro- and nanoelectronics. In this paper, we have shown that such generation can be realized as a result of the excitation by a pulsed electron source of a doublet of quasi-stationary states of a three-barrier heterostructure in the form of a symmetric double quantum well. An exciting electron pulse in the form of a Gaussian wave packet of picosecond duration, in turn, can be created, for example, by pulsed photoemission when the photocathode is exposed to a femtosecond light pulse or in some other way. The results of numerical-analytical modeling of the formation of the probability density waves and the probability current density waves outside the heterostructure are based on the solution of the nonstationary Schr?dinger equation describing the scattering of a Gaussian wave packet on a model structure formed by three tunnel-transparent dielectric films modeled by  $\delta$-barriers of the same power separated by thin conducting or vacuum nanometer layers thickness. This simplified model made it possible to implement numerical calculations and estimate the frequencies, wavelengths, and velocities of such waves, as well as the amplitudes of oscillations of probability density and current at a given intensity of the exciting packet and the power of potential barriers. The characteristics of the generated waves strongly depend on the parameters of the heterostructure. By varying the parameters of the heterostructure, one can change the energies, difference frequencies, and lifetimes of quasi-stationary doublet states. For layer thicknesses of $1 - 10^2$ nm and barrier heights of 0.5 - 2.5 eV, it is possible to provide the lifetimes of quasi-stationary states of $10^{-2} - 3\cdot10^2$ ps, the generated difference frequencies for them and the radiated waves of probability and current densities of $10^{11} - 10^{14}$ Hz, and the wavelengths of these waves $10 - 10^3$  nm. The process of emission of electron waves can be repeated or even amplified if a periodic resonant pumping of the doublet population in the heterostructure is provided by a series of Gaussian pulses with a suitable duty cycle, incident on the heterostructure in phase with oscillations of the probability and current densities.

The simple quantum mechanical model discussed in this article makes it possible to rigorously reveal the main regularities and estimate the contributions of the main characteristics and singularities to the process of excitation of waves of probability densities and currents during scattering of wave packets on a double-well heterostructure. This enables us to study in detail the properties of the complete system of wave functions of the stationary scattering problem, which forms a natural basis of unperturbed states of the zero approximation for more realistic models and methods for describing and calculating the studied generation processes. In particular, this refers to the models of fast photoemission in an open system, when, in order to describe the excitation and structure of the scattered wave packet, it is necessary to take into account the interactions of electrons with photons and with other particles in the subsequent application of the density matrix method for mixed quantum states.

%\section{Appendix}
%\subsection{Appendix}

%-----------------------------------------------------------
\appendix*
%-----------------------------------------------------------
\section{}
%\subsection{}\label{ap1}
\textbf{Appendix A} 
\vspace{0.5cm}

The essence of the method proposed by G.F. Drukarev \cite{Druk1951, Baz1969} for an analytical estimate of the contributions of singularities of integrands of the type \eqref{eq:math:8}, in short, is that after substituting \eqref{eq:math:1} into \eqref{eq:math:8} and passing to a variable  $k = \hbar ^{ - 1} \sqrt {2mE}$, each of the seven exponential terms \eqref{eq:math:1} leads to an estimated integral of the form
\begin{equation}\label{eq:math:A1}
I = \int\limits_0^\infty  {F(k)\exp ( - i\beta _t (k - k_S )^2 )dk}, 
\end{equation}
where  $\beta _t  = \hbar t/2m$, and the quantities  $k_S$ and $F(k)$  are different for the seven terms \eqref{eq:math:1}, they depend on $x,t$  and the parameters of the problem, the functions $F(k)$  can have poles $k_R$  or other singularities in the plane of the complex variable  $k$, which are determined by the features of $c_E  = c(E(k))$
  and of amplitudes of the reflected and transmitted waves. Integral \eqref{eq:math:A1} is usually estimated based on the saddle point method \cite{Lavr1967}, \cite{Peis2011}. 

\begin{figure}[h]
\includegraphics[width=8.5 cm]{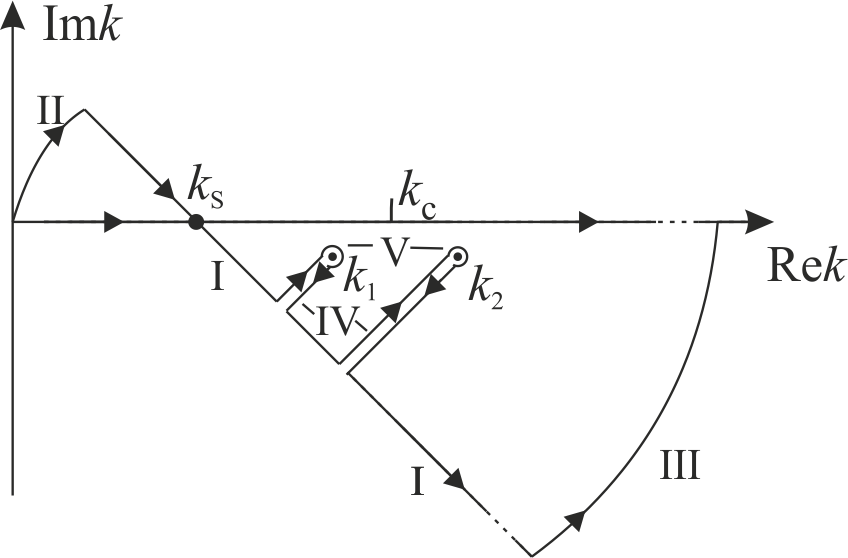}\\
  \caption{Deformation of the spectral integral contour  in the area of  analyticity  to the line I of the largest slope crossing the saddle point  $k_S$,  the poles $k_R=k_1$  and $k_R=k_2$  of the under-integral expressions are going around by small circles V.}\label{FIG:Fig3}
\end{figure}

For very large values  $\beta_t$, the main contribution to it is associated with the so-called stationary point $k=k_S$ on the real axis, which is a saddle point for a function $\operatorname{Re} ( - i\beta _t (k - k_S )^2 )$
  with respect to variables $\operatorname{Re} k$
  and  $\operatorname{Im} k$, moreover, the line of the fastest change of this function (the line of the greatest slope) is a straight line, let us denote it by I, passing on the plane of the complex variable $k$  through a point $k_S$  at an angle $-\pi/4$  to the real axis 
  (Fig.\ref{FIG:Fig3}). In accordance with the general rule, the contribution from the $k_S$  neighborhood is found by deforming the integration contour in the region of analyticity of the integrand so that it passes through the saddle point along the line I of the greatest slope. The contribution of the saddle point is usually estimated by the Poisson integral along the line I, in our case it is equal to
\begin{equation}\label{eq:math:A2}
I_{k_S }  = F(k_S )\sqrt {\frac{{ - i\pi }}
{{\beta _t }}} 
\end{equation}
the contributions of other distant parts of the deformed contour are usually small (lines II and III in Fig.\ref{FIG:Fig3}) in comparison with it.

If, when the contour is displaced near the saddle point, a pole $k_P$  or a branch point of the function $F(k)$  is encountered, then they should be bypassed along a path of type IV, V, as shown in the figure. In the case of poles, the contribution of sections IV cancels out, and the contribution of the small circle V around the pole $k_P$  is equal to the residue at this pole
\begin{equation}\label{eq:math:A3}
I_p  =  \pm 2\pi iRes\{ F(k_P )\} \exp \left( { - i\beta _t (k_P  - k_S )^2 } \right)
\end{equation}
and may not be small in comparison with the contribution of the saddle point. We take the plus sign if the pole is located in sector II of the upper half-plane, minus - in sector III of the lower half-plane, as in the figure Fig.\ref{FIG:Fig3} (by passing the pole counterclockwise or clockwise). The contributions of type \eqref{eq:math:A3} are the main ones in the ranges of values of    $t$ and  $x$, which are of interest to us, describing the oscillatory-wave behavior of the quantities $n(x,t)$  and $j(x,t)$. 

\vspace{0.5cm}

%\subsection{}\label{ap2}
\textbf{Appendix B }
\vspace{0.5cm}

Let us write down analytical formulas that describe the coordinate-time dependence of quantities $n(x,t)$   and $j(x,t)$  in the region of validity of expression \eqref{eq:math:14} for such values of $t$  and   $x$, at which the oscillatory-wave mode of beats of quasi-stationary states is established, because the main contribution is made by the pole features of the scattering amplitudes, and it is already possible to neglect small terms  $\Psi _0 (x,t)$,  $\Psi _n (x,t)$, $\Psi _3 (x,t)$  in (14).

In the regions inside each of the two wells at  $n'd \leqslant x \leqslant nd$, $n'\, = n - 1$,  $n=1,2$, substitution by the second line \eqref{eq:math:14} in \eqref{eq:math:9} and \eqref{eq:math:10} gives

\begin{widetext}
\begin{equation}\label{eq:math:B1}
\begin{gathered}
  n(x,t) = \sum\limits_{p = 1}^2 {|{\kern 1pt} \tilde B_{nE_p } |^2 } e^{2\left| {\tilde k''_p } \right|(x - x_{n-1}) - 2\left| {E''_p } \right|t}  + \sum\limits_{p = 1}^2 {|\tilde A_{nE_p } |^2 } e^{ - 2\left| {\tilde k''_p } \right|(x - x_{n-1}) - 2\left| {E''_p } \right|t}  +  \hfill \\
  \quad \quad \;\; + 2\sum\limits_{p = 1}^2 {|{\kern 1pt} \tilde A_{nE_p } \tilde B_{nE_p } |} \cos\left( {{\kern 1pt} {\kern 1pt} 2\tilde k'_p (x - x_{n-1}) + \alpha _{np}  - {\kern 1pt} {\kern 1pt} \beta _{np} } \right)e^{ - 2\left| {E''_p } \right|t}  +  \hfill \\
  \quad  + 2\left| {\tilde A_{nE_1 } \tilde A_{nE_2 } } \right|\cos\left( {\omega {\kern 1pt} {\kern 1pt} t - {\kern 1pt} {\kern 1pt} (\tilde k'_2  - \tilde k'_1 )(x - x_{n-1}) + \alpha _{n1}  - {\kern 1pt} {\kern 1pt} \alpha _{n2} } \right)e^{ - (\left| {\tilde k''_1 } \right| + \left| {\tilde k''_2 } \right|)(x - x_{n-1}) - (\left| {E''_1 } \right| + \left| {E''_2 } \right|)t}  +  \hfill \\
  \quad  + 2\left| {\tilde B_{nE_1 } \tilde B_{nE_2 } } \right|\cos\left( {\omega {\kern 1pt} {\kern 1pt} t + {\kern 1pt} {\kern 1pt} (\tilde k'_2  - \tilde k'_1 )(x - x_{n-1}) + \beta _{n1}  - {\kern 1pt} {\kern 1pt} \beta _{n2} } \right)e^{\,\;(\left| {\tilde k''_1 } \right| + \left| {\tilde k''_2 } \right|)(x - x_{n-1}) - (\left| {E''_1 } \right| + \left| {E''_2 } \right|)t} {\kern 1pt} {\kern 1pt} \, +  \hfill \\
  \quad  + 2\left| {\tilde A_{nE_1 } \tilde B_{nE_2 } } \right|\cos\left( {\omega {\kern 1pt} {\kern 1pt} t + {\kern 1pt} {\kern 1pt} (\tilde k'_2  + \tilde k'_1 )(x - x_{n-1}) + \alpha _{n1}  - {\kern 1pt} {\kern 1pt} \beta _{n2} } \right)e^{ - (\left| {\tilde k''_1 } \right| - \left| {\tilde k''_2 } \right|)(x - x_{n-1}) - (\left| {E''_1 } \right| + \left| {E''_2 } \right|)t}  +  \hfill \\
  \quad  + 2\left| {\tilde A_{nE_2 } \tilde B_{nE_1 } } \right|\cos\left( {\omega {\kern 1pt} {\kern 1pt} t - {\kern 1pt} {\kern 1pt} (\tilde k'_2  + \tilde k'_1 )(x - x_{n-1}) + \beta _{n1}  - {\kern 1pt} {\kern 1pt} \alpha _{n2} } \right)e^{\;\;(\left| {\tilde k''_1 } \right| - \left| {\tilde k''_2 } \right|)(x - x_{n-1}) - (\left| {E''_1 } \right| + \left| {E''_2 } \right|)t} {\kern 1pt} \; +  \hfill \\ 
\end{gathered} 
\end{equation}

\begin{equation}\label{eq:math:B2}
\begin{gathered}
  j(x,t) = \frac{\hbar }
{m}{\kern 1pt} \left[ {\sum\limits_{p = 1}^2 {\tilde k'_p \left( {|\tilde A_{nE_p } |^2 e^{ - 2\left| {\tilde k''_p } \right|(x - n'd)}  - \;|{\kern 1pt} \tilde B_{nE_p } |^2 e^{ + 2\left| {\tilde k''_p } \right|(x - n'd)} } \right)\;} e^{ - 2\left| {E''_p } \right|t}  - } \right. \hfill \\
  \quad \quad \quad \quad \;\; - 2\sum\limits_{p = 1}^2 {k''_p \,|{\kern 1pt} \tilde A_{nE_p } \tilde B_{nE_p } |} \;sin\left( {{\kern 1pt} {\kern 1pt} 2\tilde k'_p (x - n'd) + \;\alpha _{np}  - {\kern 1pt} {\kern 1pt} \beta _{np} } \right)e^{ - 2\left| {E''_p } \right|t}  +  \hfill \\
   + \left( {\tilde k'_2  + \tilde k'_1 } \right)\left| {\tilde A_{nE_1 } \tilde A_{nE_2 } } \right|cos\left( {\omega {\kern 1pt} {\kern 1pt} t - {\kern 1pt} {\kern 1pt} (\tilde k'_2  - \tilde k'_1 )(x - n'd) + \alpha _{n1}  - {\kern 1pt} {\kern 1pt} \alpha _{n2} } \right)e^{ - (\left| {\tilde k''_1 } \right| + \left| {\tilde k''_2 } \right|)(x - n'd) - (\left| {E''_1 } \right| + \left| {E''_2 } \right|)t} \_ \hfill \\
   - \left( {\tilde k'_2  + \tilde k'_1 } \right)\left| {\tilde B_{nE_1 } \tilde B_{nE_2 } } \right|cos\left( {\omega {\kern 1pt} {\kern 1pt} t + {\kern 1pt} {\kern 1pt} (\tilde k'_2  - \tilde k'_1 )(x - n'd) + \;\beta _{n1}  - {\kern 1pt} {\kern 1pt} \beta _{n2} } \right)e^{(\left| {\tilde k''_1 } \right| + \left| {\tilde k''_2 } \right|)(x - n'd) - (\left| {E''_1 } \right| + \left| {E''_2 } \right|)t}  +  \hfill \\
   + \left( {\tilde k'_1  - \tilde k'_2 } \right)\left| {\tilde A_{nE_1 } \tilde B_{nE_2 } } \right|cos\left( {\omega {\kern 1pt} {\kern 1pt} t + {\kern 1pt} {\kern 1pt} \left( {\tilde k'_2  + \tilde k'_1 } \right)(x - n'd) + \;\alpha _{n1}  - {\kern 1pt} {\kern 1pt} \beta _{n2} } \right)e^{ - (\left| {\tilde k''_1 } \right| - \left| {\tilde k''_2 } \right|)(x - n'd) - (\left| {E''_1 } \right| + \left| {E''_2 } \right|)t}  -  \hfill \\
  \left. { - \left( {\tilde k'_1  - \tilde k'_2 } \right)\left| {\tilde A_{nE_2 } \tilde B_{nE_1 } } \right|cos\left( {\omega {\kern 1pt} {\kern 1pt} t - {\kern 1pt} {\kern 1pt} \left( {\tilde k'_2  + \tilde k'_1 } \right)(x - n'd) + \beta _{n1}  - {\kern 1pt} \alpha _{n2} } \right)e^{(\left| {\tilde k''_1 } \right| - \left| {\tilde k''_2 } \right|)(x - n'd) - (\left| {E''_1 } \right| + \left| {E''_2 } \right|)t} } \right] \hfill \\ 
\end{gathered} 
\end{equation}
\end{widetext}
In \eqref{eq:math:B2}, we neglected small terms proportional $k'_p$  everywhere except for the second line, in which we wrote out a similar negligible sum just to illustrate the symmetry of the entire expression.

In the region to the left of the double well at  $x<0$, substitution of the first line of \eqref{eq:math:14} in \eqref{eq:math:9} and \eqref{eq:math:10} gives expressions describing damped waves traveling to the left

\begin{widetext}
\begin{equation}\label{eq:math:B3}
\begin{gathered}
  n(x,t) = \sum\limits_{p = 1}^2 {|{\kern 1pt} \tilde B_{0E_p } |^2 } e^{2\left| {k''_p } \right|x - 2\left| {E''_p } \right|t}  +  \hfill \\
  \quad \quad \quad  + 2\left| {\tilde B_{0E_1 } \tilde B_{0E_2 } } \right|cos\left( {\omega {\kern 1pt} {\kern 1pt} t + {\kern 1pt} {\kern 1pt} (k'_2  - k'_1 )x + \beta _{01}  - {\kern 1pt} {\kern 1pt} \beta _{02} } \right)e^{(\left| {k''_1 } \right| + \left| {k''_2 } \right|)x - (\left| {E''_1 } \right| + \left| {E''_2 } \right|)t} , \hfill \\ 
\end{gathered} 
\end{equation}
\begin{equation}\label{eq:math:B4}
\begin{gathered}
  j(x,t) =  - \frac{\hbar }
{m}{\kern 1pt} \left[ {\sum\limits_{p = 1}^2 {k'_p |\tilde B_{0E_p } |^2 } e^{2\left| {k''_p } \right|x - 2\left| {E''_p } \right|t} } \right. +  \hfill \\
  \quad \quad \quad  + \left. {(k'_1  + k'_2 )|\tilde B_{0E_1 } \tilde B_{0E_2 } |cos\left( {\omega {\kern 1pt} {\kern 1pt} t + {\kern 1pt} {\kern 1pt} (k'_2  - k'_1 )x + \beta _{01}  - {\kern 1pt} {\kern 1pt} \beta _{02} } \right)e^{(\left| {k''_1 } \right| + \left| {k''_2 } \right|)x - (\left| {E''_1 } \right| + \left| {E''_2 } \right|)t} } \right],\quad  \hfill \\ 
\end{gathered} 
\end{equation}
\end{widetext}

In the region to the right of the double well at  $x>x_2$, substitution of the first line of \eqref{eq:math:14} in \eqref{eq:math:9} and \eqref{eq:math:10} gives expressions describing damped waves traveling to the right

\begin{widetext}
\begin{equation}\label{eq:math:B5}
\begin{gathered}
  n(x,t) = \sum\limits_{p = 1}^2 {|\tilde A_{3E_p } ||^2 } e^{ - 2\left| {k''_p } \right|(x - x_2 ) - 2\left| {E''_p } \right|t}  +  \hfill \\
  \quad \quad  + 2\left| {\tilde A_{3E_1 } \tilde A_{3E_2 } } \right|cos\left( {\omega {\kern 1pt} {\kern 1pt} t - {\kern 1pt} {\kern 1pt} (k'_2  - k'_1 )(x - x_2 ) + \;\alpha _{31}  - {\kern 1pt} {\kern 1pt} \alpha _{32} } \right)e^{ - (\left| {k''_1 } \right| + \left| {k''_2 } \right|)(x - x_2 d) - (\left| {E''_1 } \right| + \left| {E''_2 } \right|)t} , \hfill \\ 
\end{gathered} 
\end{equation}
\begin{equation}\label{eq:math:B6}
\begin{gathered}
  j(x,t) = \frac{\hbar }
{m}{\kern 1pt} \left[ {\sum\limits_{p = 1}^2 {k'_p |\tilde A_{3E_p } |^2 } e^{ - 2\left| {k''_p } \right|(x - x_2 ) - 2\left| {E''_p } \right|t} } \right. +  \hfill \\
  \quad \quad \quad  + \left. {(k'_1  + k'_2 )\,\left| {\tilde A_{3E_1 } \tilde A_{3E_2 } } \right|cos\left( {\omega {\kern 1pt} {\kern 1pt} t - {\kern 1pt} {\kern 1pt} (k'_2  - k'_1 )(x - x_2 ) + \;\alpha _{31}  - {\kern 1pt} {\kern 1pt} \alpha _{32} } \right)e^{ - (\left| {k''_1 } \right| + \left| {k''_2 } \right|)(x - x_2 ) - (\left| {E''_1 } \right| + \left| {E''_2 } \right|)t} } \right]. \hfill \\ 
\end{gathered} 
\end{equation}
\end{widetext}

\end{document}